%
\ifx\mnmacrosloaded\undefined 
%
%
%
%

\catcode `\@=11 

\def\@version{1.4}
\def\@verdate{22nd Feb 1994}

%
%
%
%


\newif\ifprod@font

\ifx\@typeface\undefined
  \def\@typeface{Comp. Modern}\prod@fontfalse
\else
  \prod@fonttrue 
\fi

\def\newfam{\alloc@8\fam\chardef\sixt@@n} 

\ifprod@font
\font\fiverm=mtr10 at 5pt
\font\fivebf=mtbx10 at 5pt
\font\fiveit=mtti10 at 5pt
\font\fivesl=mtsl10 at 5pt
\font\fivett=mttt10 at 5pt     \hyphenchar\fivett=-1
\font\fivecsc=mtcsc10 at 5pt
\font\fivesf=mtss10 at 5pt
\font\fivei=mtmi10 at 5pt      \skewchar\fivei='177
\font\fivemib=mtmib10 at 5pt   \skewchar\fivemib='177
\font\fivesy=mtsy10 at 5pt     \skewchar\fivesy='60
\font\fivesyb=mtbsy10 at 5pt   \skewchar\fivesyb='60

\font\sixrm=mtr10 at 6pt
\font\sixbf=mtbx10 at 6pt
\font\sixit=mtti10 at 6pt
\font\sixsl=mtsl10 at 6pt
\font\sixtt=mttt10 at 6pt      \hyphenchar\sixtt=-1
\font\sixcsc=mtcsc10 at 6pt
\font\sixsf=mtss10 at 6pt
\font\sixi=mtmi10 at 6pt       \skewchar\sixi='177
\font\sixmib=mtmib10 at 6pt    \skewchar\sixmib='177
\font\sixsy=mtsy10 at 6pt      \skewchar\sixsy='60
\font\sixsyb=mtbsy10 at 6pt    \skewchar\sixsyb='60

\font\sevenrm=mtr10 at 7pt
\font\sevenbf=mtbx10 at 7pt
\font\sevenit=mtti10 at 7pt
\font\sevensl=mtsl10 at 7pt
\font\seventt=mttt10 at 7pt     \hyphenchar\seventt=-1
\font\sevencsc=mtcsc10 at 7pt
\font\sevensf=mtss10 at 7pt
\font\seveni=mtmi10 at 7pt      \skewchar\seveni='177
\font\sevenmib=mtmib10 at 7pt   \skewchar\sevenmib='177
\font\sevensy=mtsy10 at 7pt     \skewchar\sevensy='60
\font\sevensyb=mtbsy10 at 7pt   \skewchar\sevensyb='60

\font\eightrm=mtr10 at 8pt
\font\eightbf=mtbx10 at 8pt
\font\eightit=mtti10 at 8pt
\font\eighti=mtmi10 at 8pt      \skewchar\eighti='177
\font\eightmib=mtmib10 at 8pt   \skewchar\eightmib='177
\font\eightsy=mtsy10 at 8pt     \skewchar\eightsy='60
\font\eightsyb=mtbsy10 at 8pt   \skewchar\eightsyb='60
\font\eightsl=mtsl10 at 8pt
\font\eighttt=mttt10 at 8pt     \hyphenchar\eighttt=-1
\font\eightcsc=mtcsc10 at 8pt
\font\eightsf=mtss10 at 8pt

\font\ninerm=mtr10 at 9pt
\font\ninebf=mtbx10 at 9pt
\font\nineit=mtti10 at 9pt
\font\ninei=mtmi10 at 9pt      \skewchar\ninei='177
\font\ninemib=mtmib10 at 9pt   \skewchar\ninemib='177
\font\ninesy=mtsy10 at 9pt     \skewchar\ninesy='60
\font\ninesyb=mtbsy10 at 9pt   \skewchar\ninesyb='60
\font\ninesl=mtsl10 at 9pt
\font\ninett=mttt10 at 9pt     \hyphenchar\ninett=-1
\font\ninecsc=mtcsc10 at 9pt
\font\ninesf=mtss10 at 9pt

\font\tenrm=mtr10
\font\tenbf=mtbx10
\font\tenit=mtti10
\font\teni=mtmi10		\skewchar\teni='177
\font\tenmib=mtmib10	\skewchar\tenmib='177
\font\tensy=mtsy10		\skewchar\tensy='60
\font\tensyb=mtbsy10	\skewchar\tensyb='60
\font\tenex=cmex10
\font\tensl=mtsl10
\font\tentt=mttt10		\hyphenchar\tentt=-1
\font\tencsc=mtcsc10
\font\tensf=mtss10

\font\elevenrm=mtr10 at 11pt
\font\elevenbf=mtbx10 at 11pt
\font\elevenit=mtti10 at 11pt
\font\eleveni=mtmi10 at 11pt      \skewchar\eleveni='177
\font\elevenmib=mtmib10 at 11pt   \skewchar\elevenmib='177
\font\elevensy=mtsy10 at 11pt     \skewchar\elevensy='60
\font\elevensyb=mtbsy10 at 11pt   \skewchar\elevensyb='60
\font\elevensl=mtsl10 at 11pt
\font\eleventt=mttt10 at 11pt     \hyphenchar\eleventt=-1
\font\elevencsc=mtcsc10 at 11pt
\font\elevensf=mtss10 at 11pt

\font\twelverm=mtr10 at 12pt
\font\twelvebf=mtbx10 at 12pt
\font\twelveit=mtti10 at 12pt
\font\twelvesl=mtsl10 at 12pt
\font\twelvett=mttt10 at 12pt     \hyphenchar\twelvett=-1
\font\twelvecsc=mtcsc10 at 12pt
\font\twelvesf=mtss10 at 12pt
\font\twelvei=mtmi10 at 12pt      \skewchar\twelvei='177
\font\twelvemib=mtmib10 at 12pt   \skewchar\twelvemib='177
\font\twelvesy=mtsy10 at 12pt     \skewchar\twelvesy='60
\font\twelvesyb=mtbsy10 at 12pt   \skewchar\twelvesyb='60

\font\fourteenrm=mtr10 at 14pt
\font\fourteenbf=mtbx10 at 14pt
\font\fourteenit=mtti10 at 14pt
\font\fourteeni=mtmi10 at 14pt      \skewchar\fourteeni='177
\font\fourteenmib=mtmib10 at 14pt   \skewchar\fourteenmib='177
\font\fourteensy=mtsy10 at 14pt     \skewchar\fourteensy='60
\font\fourteensyb=mtbsy10 at 14pt   \skewchar\fourteensyb='60
\font\fourteensl=mtsl10 at 14pt
\font\fourteentt=mttt10 at 14pt     \hyphenchar\fourteentt=-1
\font\fourteencsc=mtcsc10 at 14pt
\font\fourteensf=mtss10 at 14pt

\font\seventeenrm=mtr10 at 17pt
\font\seventeenbf=mtbx10 at 17pt
\font\seventeenit=mtti10 at 17pt
\font\seventeeni=mtmi10 at 17pt      \skewchar\seventeeni='177
\font\seventeenmib=mtmib10 at 17pt   \skewchar\seventeenmib='177
\font\seventeensy=mtsy10 at 17pt     \skewchar\seventeensy='60
\font\seventeensyb=mtbsy10 at 17pt   \skewchar\seventeensyb='60
\font\seventeensl=mtsl10 at 17pt
\font\seventeentt=mttt10 at 17pt     \hyphenchar\seventeentt=-1
\font\seventeencsc=mtcsc10 at 17pt
\font\seventeensf=mtss10 at 17pt


\newfam\xmfam
\newfam\ymfam

\font\fivexm=mtxm10 at 5pt
\font\sixxm=mtxm10 at 6pt
\font\sevenxm=mtxm10 at 7pt
\font\eightxm=mtxm10 at 8pt
\font\ninexm=mtxm10 at 9pt
\font\tenxm=mtxm10
\font\elevenxm=mtxm10 at 11pt
\font\twelvexm=mtxm10 at 12pt
\font\fourteenxm=mtxm10 at 14pt
\font\seventeenxm=mtxm10 at 17pt

\font\fiveym=mtym10 at 5pt
\font\sixym=mtym10 at 6pt
\font\sevenym=mtym10 at 7pt
\font\eightym=mtym10 at 8pt
\font\nineym=mtym10 at 9pt
\font\tenym=mtym10
\font\elevenym=mtym10 at 11pt
\font\twelveym=mtym10 at 12pt
\font\fourteenym=mtym10 at 14pt
\font\seventeenym=mtym10 at 17pt
\else
\font\fiverm=cmr5
\font\fivei=cmmi5             \skewchar\fivei='177
\font\fivemib=cmmib10 at 5pt  \skewchar\fivemib='177
\font\fivesy=cmsy5            \skewchar\fivesy='60
\font\fivesyb=cmbsy10 at 5pt  \skewchar\fivesyb='60
\font\fivebf=cmbx5

\font\sixrm=cmr6
\font\sixi=cmmi6             \skewchar\sixi='177
\font\sixmib=cmmib10 at 6pt  \skewchar\sixmib='177
\font\sixsy=cmsy6            \skewchar\sixsy='60
\font\sixsyb=cmbsy10 at 6pt  \skewchar\sixsyb='60
\font\sixbf=cmbx6

\font\sevenrm=cmr7
\font\seveni=cmmi7             \skewchar\seveni='177
\font\sevenmib=cmmib10 at 7pt  \skewchar\sevenmib='177
\font\sevensy=cmsy7            \skewchar\sevensy='60
\font\sevensyb=cmbsy10 at 7pt  \skewchar\sevensyb='60
\font\sevenbf=cmbx7

\font\eightrm=cmr8
\font\eightbf=cmbx8
\font\eightit=cmti8
\font\eighti=cmmi8			\skewchar\eighti='177
\font\eightmib=cmmib10 at 8pt	\skewchar\eightmib='177
\font\eightsy=cmsy8			\skewchar\eightsy='60
\font\eightsyb=cmbsy10 at 8pt	\skewchar\eightsyb='60
\font\eightsl=cmsl8
\font\eighttt=cmtt8			\hyphenchar\eighttt=-1
\font\eightcsc=cmcsc10 at 8pt
\font\eightsf=cmss8

\font\ninerm=cmr9
\font\ninebf=cmbx9
\font\nineit=cmti9
\font\ninei=cmmi9			\skewchar\ninei='177
\font\ninemib=cmmib10 at 9pt	\skewchar\ninemib='177
\font\ninesy=cmsy9			\skewchar\ninesy='60
\font\ninesyb=cmbsy10 at 9pt	\skewchar\ninesyb='60
\font\ninesl=cmsl9
\font\ninett=cmtt9			\hyphenchar\ninett=-1
\font\ninecsc=cmcsc10 at 9pt
\font\ninesf=cmss9

\font\tenrm=cmr10
\font\tenbf=cmbx10
\font\tenit=cmti10
\font\teni=cmmi10		\skewchar\teni='177
\font\tenmib=cmmib10	\skewchar\tenmib='177
\font\tensy=cmsy10		\skewchar\tensy='60
\font\tensyb=cmbsy10	\skewchar\tensyb='60
\font\tenex=cmex10
\font\tensl=cmsl10
\font\tentt=cmtt10		\hyphenchar\tentt=-1
\font\tencsc=cmcsc10
\font\tensf=cmss10

\font\elevenrm=cmr10 scaled \magstephalf
\font\elevenbf=cmbx10 scaled \magstephalf
\font\elevenit=cmti10 scaled \magstephalf
\font\eleveni=cmmi10 scaled \magstephalf	\skewchar\eleveni='177
\font\elevenmib=cmmib10 scaled \magstephalf	\skewchar\elevenmib='177
\font\elevensy=cmsy10 scaled \magstephalf	\skewchar\elevensy='60
\font\elevensyb=cmbsy10 scaled \magstephalf	\skewchar\elevensyb='60
\font\elevensl=cmsl10 scaled \magstephalf
\font\eleventt=cmtt10 scaled \magstephalf	\hyphenchar\eleventt=-1
\font\elevencsc=cmcsc10 scaled \magstephalf
\font\elevensf=cmss10 scaled \magstephalf

\font\twelverm=cmr10 scaled \magstep1
\font\twelvebf=cmbx10 scaled \magstep1
\font\twelvei=cmmi10 scaled \magstep1      \skewchar\twelvei='177
\font\twelvemib=cmmib10 scaled \magstep1   \skewchar\twelvemib='177
\font\twelvesy=cmsy10 scaled \magstep1     \skewchar\twelvesy='60
\font\twelvesyb=cmbsy10 scaled \magstep1   \skewchar\twelvesyb='60

\font\fourteenrm=cmr10 scaled \magstep2
\font\fourteenbf=cmbx10 scaled \magstep2
\font\fourteenit=cmti10 scaled \magstep2
\font\fourteeni=cmmi10 scaled \magstep2		\skewchar\fourteeni='177
\font\fourteenmib=cmmib10 scaled \magstep2	\skewchar\fourteenmib='177
\font\fourteensy=cmsy10 scaled \magstep2	\skewchar\fourteensy='60
\font\fourteensyb=cmbsy10 scaled \magstep2	\skewchar\fourteensyb='60
\font\fourteensl=cmsl10 scaled \magstep2
\font\fourteentt=cmtt10 scaled \magstep2	\hyphenchar\fourteentt=-1
\font\fourteencsc=cmcsc10 scaled \magstep2
\font\fourteensf=cmss10 scaled \magstep2

\font\seventeenrm=cmr10 scaled \magstep3
\font\seventeenbf=cmbx10 scaled \magstep3
\font\seventeenit=cmti10 scaled \magstep3
\font\seventeeni=cmmi10 scaled \magstep3	\skewchar\seventeeni='177
\font\seventeenmib=cmmib10 scaled \magstep3	\skewchar\seventeenmib='177
\font\seventeensy=cmsy10 scaled \magstep3	\skewchar\seventeensy='60
\font\seventeensyb=cmbsy10 scaled \magstep3	\skewchar\seventeensyb='60
\font\seventeensl=cmsl10 scaled \magstep3
\font\seventeentt=cmtt10 scaled \magstep3	\hyphenchar\seventeentt=-1
\font\seventeencsc=cmcsc10 scaled \magstep3
\font\seventeensf=cmss10 scaled \magstep3
\fi

\def\hexnumber#1{\ifcase#1 0\or1\or2\or3\or4\or5\or6\or7\or8\or9\or
  A\or B\or C\or D\or E\or F\fi}

\ifprod@font
  \edef\@xm{\hexnumber\xmfam}
  \edef\@ym{\hexnumber\ymfam}
\fi

\def\makestrut{%
  \setbox\strutbox=\hbox{%
    \vrule height.7\baselineskip depth.3\baselineskip width \z@}%
}

\def\baselinestretch{1}
\newskip\tmp@bls

\def\b@ls#1{
  \tmp@bls=#1\relax
  \baselineskip=#1\relax\makestrut
  \normalbaselineskip=\baselinestretch\tmp@bls
  \normalbaselines
}

\def\nostb@ls#1{
  \normalbaselineskip=#1\relax
  \normalbaselines
  \makestrut
}

%

\newfam\mibfam 
\newfam\sybfam 
\newfam\scfam  
\newfam\sffam  

\def\mit{\fam\@ne}

\def\cal{\fam\tw@}

\def\em{\ifdim\fontdimen1\font>\z@ \rm\else\it\fi}

\textfont3=\tenex
\scriptfont3=\tenex
\scriptscriptfont3=\tenex

\setbox0=\hbox{\tenex B} \p@renwd=\wd0 

\def\eightpoint{
  \def\rm{\fam0\eightrm}%
  \textfont0=\eightrm \scriptfont0=\sixrm \scriptscriptfont0=\fiverm%
  \textfont1=\eighti  \scriptfont1=\sixi  \scriptscriptfont1=\fivei%
  \textfont2=\eightsy \scriptfont2=\sixsy \scriptscriptfont2=\fivesy%
  \textfont\itfam=\eightit\def\it{\fam\itfam\eightit}%
  \ifprod@font
    \scriptfont\itfam=\sixit
      \scriptscriptfont\itfam=\fiveit
  \else
    \scriptfont\itfam=\eightit
      \scriptscriptfont\itfam=\eightit
  \fi
  \textfont\bffam=\eightbf%
    \scriptfont\bffam=\sixbf%
      \scriptscriptfont\bffam=\fivebf%
  \def\bf{\fam\bffam\eightbf}%
  \textfont\slfam=\eightsl\def\sl{\fam\slfam\eightsl}%
  \ifprod@font
    \scriptfont\slfam=\sixsl
      \scriptscriptfont\slfam=\fivesl
  \else
    \scriptfont\slfam=\eightsl
      \scriptscriptfont\slfam=\eightsl
  \fi
  \textfont\ttfam=\eighttt\def\tt{\fam\ttfam\eighttt}%
  \ifprod@font
    \scriptfont\ttfam=\sixtt
      \scriptscriptfont\ttfam=\fivett
  \else
    \scriptfont\ttfam=\eighttt
      \scriptscriptfont\ttfam=\eighttt
  \fi
  \textfont\scfam=\eightcsc\def\sc{\fam\scfam\eightcsc}%
  \ifprod@font
    \scriptfont\scfam=\sixcsc
      \scriptscriptfont\scfam=\fivecsc
  \else
    \scriptfont\scfam=\eightcsc
      \scriptscriptfont\scfam=\eightcsc
  \fi
  \textfont\sffam=\eightsf\def\sf{\fam\sffam\eightsf}%
  \ifprod@font
    \scriptfont\sffam=\sixsf
      \scriptscriptfont\sffam=\fivesf
  \else
    \scriptfont\sffam=\eightsf
      \scriptscriptfont\sffam=\eightsf
  \fi
  \textfont\mibfam=\eightmib
    \scriptfont\mibfam=\sixmib
      \scriptscriptfont\mibfam=\fivemib
  \textfont\sybfam=\eightsyb
    \scriptfont\sybfam=\sixsyb
      \scriptscriptfont\sybfam=\fivesyb
  \ifprod@font
    \textfont\xmfam=\eightxm
      \scriptfont\xmfam=\sixxm
        \scriptscriptfont\xmfam=\fivexm
    \textfont\ymfam=\eightym
      \scriptfont\ymfam=\sixym
        \scriptscriptfont\ymfam=\fiveym
  \fi
  \def\oldstyle{\fam\@ne\eighti}%
  \def\boldstyle{\fam\mibfam\eightmib}%
  \b@ls{10pt}\rm%
}

\def\ninepoint{
  \def\rm{\fam0\ninerm}%
  \textfont0=\ninerm \scriptfont0=\sixrm \scriptscriptfont0=\fiverm%
  \textfont1=\ninei  \scriptfont1=\sixi  \scriptscriptfont1=\fivei%
  \textfont2=\ninesy \scriptfont2=\sixsy \scriptscriptfont2=\fivesy%
  \textfont\itfam=\nineit\def\it{\fam\itfam\nineit}%
  \ifprod@font
    \scriptfont\itfam=\sixit
      \scriptscriptfont\itfam=\fiveit
  \else
    \scriptfont\itfam=\nineit
      \scriptscriptfont\itfam=\nineit
  \fi
  \textfont\bffam=\ninebf%
    \scriptfont\bffam=\sixbf%
      \scriptscriptfont\bffam=\fivebf%
  \def\bf{\fam\bffam\ninebf}%
  \textfont\slfam=\ninesl\def\sl{\fam\slfam\ninesl}%
  \ifprod@font
    \scriptfont\slfam=\sixsl
      \scriptscriptfont\slfam=\fivesl
  \else
    \scriptfont\slfam=\ninesl
      \scriptscriptfont\slfam=\ninesl
  \fi
  \textfont\ttfam=\ninett\def\tt{\fam\ttfam\ninett}%
  \ifprod@font
    \scriptfont\ttfam=\sixtt
      \scriptscriptfont\ttfam=\fivett
  \else
    \scriptfont\ttfam=\ninett
      \scriptscriptfont\ttfam=\ninett
  \fi
  \textfont\scfam=\ninecsc\def\sc{\fam\scfam\ninecsc}%
  \ifprod@font
    \scriptfont\scfam=\sixcsc
      \scriptscriptfont\scfam=\fivecsc
  \else
    \scriptfont\scfam=\ninecsc
      \scriptscriptfont\scfam=\ninecsc
  \fi
  \textfont\sffam=\ninesf\def\sf{\fam\sffam\ninesf}%
  \ifprod@font
    \scriptfont\sffam=\sixsf
      \scriptscriptfont\sffam=\fivesf
  \else
    \scriptfont\sffam=\ninesf
      \scriptscriptfont\sffam=\ninesf
  \fi
  \textfont\mibfam=\ninemib
    \scriptfont\mibfam=\sixmib
      \scriptscriptfont\mibfam=\fivemib
  \textfont\sybfam=\ninesyb
    \scriptfont\sybfam=\sixsyb
      \scriptscriptfont\sybfam=\fivesyb
  \ifprod@font
    \textfont\xmfam=\ninexm
      \scriptfont\xmfam=\sixxm
        \scriptscriptfont\xmfam=\fivexm
    \textfont\ymfam=\nineym
      \scriptfont\ymfam=\sixym
        \scriptscriptfont\ymfam=\fiveym
  \fi
  \def\oldstyle{\fam\@ne\ninei}%
  \def\boldstyle{\fam\mibfam\ninemib}%
  \b@ls{\TextLeading plus \Feathering}\rm%
}

\def\tenpoint{
  \def\rm{\fam0\tenrm}%
  \textfont0=\tenrm \scriptfont0=\sevenrm \scriptscriptfont0=\fiverm%
  \textfont1=\teni  \scriptfont1=\seveni  \scriptscriptfont1=\fivei%
  \textfont2=\tensy \scriptfont2=\sevensy \scriptscriptfont2=\fivesy%
  \textfont\itfam=\tenit\def\it{\fam\itfam\tenit}%
  \ifprod@font
    \scriptfont\itfam=\sevenit
      \scriptscriptfont\itfam=\fiveit
  \else
    \scriptfont\itfam=\tenit
      \scriptscriptfont\itfam=\tenit
  \fi
  \textfont\bffam=\tenbf%
    \scriptfont\bffam=\sevenbf%
      \scriptscriptfont\bffam=\fivebf%
  \def\bf{\fam\bffam\tenbf}%
  \textfont\slfam=\tensl\def\sl{\fam\slfam\tensl}%
  \ifprod@font
    \scriptfont\slfam=\sevensl
      \scriptscriptfont\slfam=\fivesl
  \else
    \scriptfont\slfam=\tensl
      \scriptscriptfont\slfam=\tensl
  \fi
  \textfont\ttfam=\tentt\def\tt{\fam\ttfam\tentt}%
  \ifprod@font
    \scriptfont\ttfam=\seventt
      \scriptscriptfont\ttfam=\fivett
  \else
    \scriptfont\ttfam=\tentt
      \scriptscriptfont\ttfam=\tentt
  \fi
  \textfont\scfam=\tencsc\def\sc{\fam\scfam\tencsc}%
  \ifprod@font
    \scriptfont\scfam=\sevencsc
      \scriptscriptfont\scfam=\fivecsc
  \else
    \scriptfont\scfam=\tencsc
      \scriptscriptfont\scfam=\tencsc
  \fi
  \textfont\sffam=\tensf\def\sf{\fam\sffam\tensf}%
  \ifprod@font
    \scriptfont\sffam=\sevensf
      \scriptscriptfont\sffam=\fivesf
  \else
    \scriptfont\sffam=\tensf
      \scriptscriptfont\sffam=\tensf
  \fi
  \textfont\mibfam=\tenmib
    \scriptfont\mibfam=\sevenmib
      \scriptscriptfont\mibfam=\fivemib
  \textfont\sybfam=\tensyb
    \scriptfont\sybfam=\sevensyb
      \scriptscriptfont\sybfam=\fivesyb
  \ifprod@font
    \textfont\xmfam=\tenxm
      \scriptfont\xmfam=\sevenxm
        \scriptscriptfont\xmfam=\fivexm
    \textfont\ymfam=\tenym
      \scriptfont\ymfam=\sevenym
        \scriptscriptfont\ymfam=\fiveym
  \fi
  \def\oldstyle{\fam\@ne\teni}%
  \def\boldstyle{\fam\mibfam\tenmib}%
  \b@ls{11pt}\rm%
}

\def\elevenpoint{
  \def\rm{\fam0\elevenrm}%
  \textfont0=\elevenrm \scriptfont0=\eightrm \scriptscriptfont0=\sixrm%
  \textfont1=\eleveni  \scriptfont1=\eighti  \scriptscriptfont1=\sixi%
  \textfont2=\elevensy \scriptfont2=\eightsy \scriptscriptfont2=\sixsy%
  \textfont\itfam=\elevenit\def\it{\fam\itfam\elevenit}%
  \ifprod@font
    \scriptfont\itfam=\eightit
      \scriptscriptfont\itfam=\sixit
  \else
    \scriptfont\itfam=\elevenit
      \scriptscriptfont\itfam=\elevenit
  \fi
  \textfont\bffam=\elevenbf%
    \scriptfont\bffam=\eightbf%
      \scriptscriptfont\bffam=\sixbf%
  \def\bf{\fam\bffam\elevenbf}%
  \textfont\slfam=\elevensl\def\sl{\fam\slfam\elevensl}%
  \ifprod@font
    \scriptfont\slfam=\eightsl
      \scriptscriptfont\slfam=\sixsl
  \else
    \scriptfont\slfam=\elevensl
      \scriptscriptfont\slfam=\elevensl
  \fi
  \textfont\ttfam=\eleventt\def\tt{\fam\ttfam\eleventt}%
  \ifprod@font
    \scriptfont\ttfam=\eighttt
      \scriptscriptfont\ttfam=\sixtt
  \else
    \scriptfont\ttfam=\eleventt
      \scriptscriptfont\ttfam=\eleventt
  \fi
  \textfont\scfam=\elevencsc\def\sc{\fam\scfam\elevencsc}%
  \ifprod@font
    \scriptfont\scfam=\eightcsc
      \scriptscriptfont\scfam=\sixcsc
  \else
    \scriptfont\scfam=\elevencsc
      \scriptscriptfont\scfam=\elevencsc
  \fi
  \textfont\sffam=\elevensf\def\sf{\fam\sffam\elevensf}%
  \ifprod@font
    \scriptfont\sffam=\eightsf
      \scriptscriptfont\sffam=\sixsf
  \else
    \scriptfont\sffam=\elevensf
      \scriptscriptfont\sffam=\elevensf
  \fi
  \textfont\mibfam=\elevenmib
    \scriptfont\mibfam=\eightmib
      \scriptscriptfont\mibfam=\sixmib
  \textfont\sybfam=\elevensyb
    \scriptfont\sybfam=\eightsyb
      \scriptscriptfont\sybfam=\sixsyb
  \ifprod@font
    \textfont\xmfam=\elevenxm
      \scriptfont\xmfam=\eightxm
       \scriptscriptfont\xmfam=\sixxm
    \textfont\ymfam=\elevenym
      \scriptfont\ymfam=\eightym
        \scriptscriptfont\ymfam=\sixym
   \fi
  \def\oldstyle{\fam\@ne\eleveni}%
  \def\boldstyle{\fam\mibfam\elevenmib}%
  \b@ls{13pt}\rm%
}

\def\fourteenpoint{
  \def\rm{\fam0\fourteenrm}%
  \textfont0\fourteenrm  \scriptfont0\tenrm  \scriptscriptfont0\sevenrm%
  \textfont1\fourteeni   \scriptfont1\teni   \scriptscriptfont1\seveni%
  \textfont2\fourteensy  \scriptfont2\tensy  \scriptscriptfont2\sevensy%
  \textfont\itfam=\fourteenit\def\it{\fam\itfam\fourteenit}%
  \ifprod@font
    \scriptfont\itfam=\tenit
      \scriptscriptfont\itfam=\sevenit
  \else
    \scriptfont\itfam=\fourteenit
      \scriptscriptfont\itfam=\fourteenit
  \fi
  \textfont\bffam=\fourteenbf%
    \scriptfont\bffam=\tenbf%
      \scriptscriptfont\bffam=\sevenbf%
  \def\bf{\fam\bffam\fourteenbf}%
  \textfont\slfam=\fourteensl\def\sl{\fam\slfam\fourteensl}%
  \ifprod@font
    \scriptfont\slfam=\tensl
      \scriptscriptfont\slfam=\sevensl
  \else
    \scriptfont\slfam=\fourteensl
      \scriptscriptfont\slfam=\fourteensl
  \fi
  \textfont\ttfam=\fourteentt\def\tt{\fam\ttfam\fourteentt}%
  \ifprod@font
    \scriptfont\ttfam=\tentt
      \scriptscriptfont\ttfam=\seventt
  \else
    \scriptfont\ttfam=\fourteentt
      \scriptscriptfont\ttfam=\fourteentt
  \fi
  \textfont\scfam=\fourteencsc\def\sc{\fam\scfam\fourteencsc}%
  \ifprod@font
    \scriptfont\scfam=\tencsc
      \scriptscriptfont\scfam=\sevencsc
  \else
    \scriptfont\scfam=\fourteencsc
      \scriptscriptfont\scfam=\fourteencsc
  \fi
  \textfont\sffam=\fourteensf\def\sf{\fam\sffam\fourteensf}%
  \ifprod@font
    \scriptfont\sffam=\tensf
      \scriptscriptfont\sffam=\sevensf
  \else
    \scriptfont\sffam=\fourteensf
      \scriptscriptfont\sffam=\fourteensf
  \fi
  \textfont\mibfam=\fourteenmib
    \scriptfont\mibfam=\tenmib
      \scriptscriptfont\mibfam=\sevenmib
  \textfont\sybfam=\fourteensyb
    \scriptfont\sybfam=\tensyb
      \scriptscriptfont\sybfam=\sevensyb
  \ifprod@font
    \textfont\xmfam=\fourteenxm
      \scriptfont\xmfam=\tenxm
        \scriptscriptfont\xmfam=\sevenxm
   \textfont\ymfam=\fourteenym
      \scriptfont\ymfam=\tenym
        \scriptscriptfont\ymfam=\sevenym
  \fi
  \def\oldstyle{\fam\@ne\fourteeni}%
  \def\boldstyle{\fam\mibfam\fourteenmib}%
  \b@ls{17pt}\rm%
}

\def\seventeenpoint{
  \def\rm{\fam0\seventeenrm}%
  \textfont0\seventeenrm  \scriptfont0\twelverm  \scriptscriptfont0\tenrm%
  \textfont1\seventeeni   \scriptfont1\twelvei   \scriptscriptfont1\teni%
  \textfont2\seventeensy  \scriptfont2\twelvesy  \scriptscriptfont2\tensy%
  \textfont\itfam=\seventeenit\def\it{\fam\itfam\seventeenit}%
  \ifprod@font
    \scriptfont\itfam=\twelveit
      \scriptscriptfont\itfam=\tenit
  \else
    \scriptfont\itfam=\seventeenit
      \scriptscriptfont\itfam=\seventeenit
  \fi
  \textfont\bffam=\seventeenbf%
    \scriptfont\bffam=\twelvebf%
      \scriptscriptfont\bffam=\tenbf%
  \def\bf{\fam\bffam\seventeenbf}%
  \textfont\slfam=\seventeensl\def\sl{\fam\slfam\seventeensl}%
  \ifprod@font
    \scriptfont\slfam=\twelvesl
      \scriptscriptfont\slfam=\tensl
  \else
    \scriptfont\slfam=\seventeensl
      \scriptscriptfont\slfam=\seventeensl
  \fi
  \textfont\ttfam=\seventeentt\def\tt{\fam\ttfam\seventeentt}%
  \ifprod@font
    \scriptfont\ttfam=\twelvett
      \scriptscriptfont\ttfam=\tentt
  \else
    \scriptfont\ttfam=\seventeentt
      \scriptscriptfont\ttfam=\seventeentt
  \fi
  \textfont\scfam=\seventeencsc\def\sc{\fam\scfam\seventeencsc}%
  \ifprod@font
    \scriptfont\scfam=\twelvecsc
      \scriptscriptfont\scfam=\tencsc
  \else
    \scriptfont\scfam=\seventeencsc
      \scriptscriptfont\scfam=\seventeencsc
  \fi
  \textfont\sffam=\seventeensf\def\sf{\fam\sffam\seventeensf}%
  \ifprod@font
    \scriptfont\sffam=\twelvesf
      \scriptscriptfont\sffam=\tensf
  \else
    \scriptfont\sffam=\seventeensf
      \scriptscriptfont\sffam=\seventeensf
  \fi
  \textfont\mibfam=\seventeenmib
    \scriptfont\mibfam=\twelvemib
      \scriptscriptfont\mibfam=\tenmib
  \textfont\sybfam=\seventeensyb
    \scriptfont\sybfam=\twelvesyb
      \scriptscriptfont\sybfam=\tensyb
  \ifprod@font
    \textfont\xmfam=\seventeenxm
      \scriptfont\xmfam=\twelvexm
        \scriptscriptfont\xmfam=\tenxm
    \textfont\ymfam=\seventeenym
      \scriptfont\ymfam=\twelveym
        \scriptscriptfont\ymfam=\tenym
  \fi
  \def\oldstyle{\fam\@ne\seventeeni}%
  \def\boldstyle{\fam\mibfam\seventeenmib}%
  \b@ls{20pt}\rm%
}

\lineskip=1pt      \normallineskip=\lineskip
\lineskiplimit=\z@ \normallineskiplimit=\lineskiplimit



\def\la{\mathrel{\mathchoice {\vcenter{\offinterlineskip\halign{\hfil
$\displaystyle##$\hfil\cr<\cr\sim\cr}}}
{\vcenter{\offinterlineskip\halign{\hfil$\textstyle##$\hfil\cr
<\cr\sim\cr}}}
{\vcenter{\offinterlineskip\halign{\hfil$\scriptstyle##$\hfil\cr
<\cr\sim\cr}}}
{\vcenter{\offinterlineskip\halign{\hfil$\scriptscriptstyle##$\hfil\cr
<\cr\sim\cr}}}}}

\def\ga{\mathrel{\mathchoice {\vcenter{\offinterlineskip\halign{\hfil
$\displaystyle##$\hfil\cr>\cr\sim\cr}}}
{\vcenter{\offinterlineskip\halign{\hfil$\textstyle##$\hfil\cr
>\cr\sim\cr}}}
{\vcenter{\offinterlineskip\halign{\hfil$\scriptstyle##$\hfil\cr
>\cr\sim\cr}}}
{\vcenter{\offinterlineskip\halign{\hfil$\scriptscriptstyle##$\hfil\cr
>\cr\sim\cr}}}}}

\def\getsto{\mathrel{\mathchoice {\vcenter{\offinterlineskip
\halign{\hfil
$\displaystyle##$\hfil\cr\gets\cr\to\cr}}}
{\vcenter{\offinterlineskip\halign{\hfil$\textstyle##$\hfil\cr\gets
\cr\to\cr}}}
{\vcenter{\offinterlineskip\halign{\hfil$\scriptstyle##$\hfil\cr\gets
\cr\to\cr}}}
{\vcenter{\offinterlineskip\halign{\hfil$\scriptscriptstyle##$\hfil\cr
\gets\cr\to\cr}}}}}

\def\lid{\mathrel{\mathchoice {\vcenter{\offinterlineskip\halign{\hfil
$\displaystyle##$\hfil\cr<\cr\noalign{\vskip1.2pt}=\cr}}}
{\vcenter{\offinterlineskip\halign{\hfil$\textstyle##$\hfil\cr<\cr
\noalign{\vskip1.2pt}=\cr}}}
{\vcenter{\offinterlineskip\halign{\hfil$\scriptstyle##$\hfil\cr<\cr
\noalign{\vskip1pt}=\cr}}}
{\vcenter{\offinterlineskip\halign{\hfil$\scriptscriptstyle##$\hfil\cr
<\cr
\noalign{\vskip0.9pt}=\cr}}}}}

\def\gid{\mathrel{\mathchoice {\vcenter{\offinterlineskip\halign{\hfil
$\displaystyle##$\hfil\cr>\cr\noalign{\vskip1.2pt}=\cr}}}
{\vcenter{\offinterlineskip\halign{\hfil$\textstyle##$\hfil\cr>\cr
\noalign{\vskip1.2pt}=\cr}}}
{\vcenter{\offinterlineskip\halign{\hfil$\scriptstyle##$\hfil\cr>\cr
\noalign{\vskip1pt}=\cr}}}
{\vcenter{\offinterlineskip\halign{\hfil$\scriptscriptstyle##$\hfil\cr
>\cr
\noalign{\vskip0.9pt}=\cr}}}}}

\def\grole{\mathrel{\mathchoice {\vcenter{\offinterlineskip\halign{\hfil
$\displaystyle##$\hfil\cr>\cr\noalign{\vskip-1.5pt}<\cr}}}
{\vcenter{\offinterlineskip\halign{\hfil$\textstyle##$\hfil\cr
>\cr\noalign{\vskip-1.5pt}<\cr}}}
{\vcenter{\offinterlineskip\halign{\hfil$\scriptstyle##$\hfil\cr
>\cr\noalign{\vskip-1pt}<\cr}}}
{\vcenter{\offinterlineskip\halign{\hfil$\scriptscriptstyle##$\hfil\cr
>\cr\noalign{\vskip-0.5pt}<\cr}}}}}

\def\leogr{\mathrel{\mathchoice {\vcenter{\offinterlineskip\halign{\hfil
$\displaystyle##$\hfil\cr<\cr\noalign{\vskip-1.5pt}>\cr}}}
{\vcenter{\offinterlineskip\halign{\hfil$\textstyle##$\hfil\cr
<\cr\noalign{\vskip-1.5pt}>\cr}}}
{\vcenter{\offinterlineskip\halign{\hfil$\scriptstyle##$\hfil\cr
<\cr\noalign{\vskip-1pt}>\cr}}}
{\vcenter{\offinterlineskip\halign{\hfil$\scriptscriptstyle##$\hfil\cr
<\cr\noalign{\vskip-0.5pt}>\cr}}}}}

\def\loa{\mathrel{\mathchoice {\vcenter{\offinterlineskip\halign{\hfil
$\displaystyle##$\hfil\cr<\cr\approx\cr}}}
{\vcenter{\offinterlineskip\halign{\hfil$\textstyle##$\hfil\cr
<\cr\approx\cr}}}
{\vcenter{\offinterlineskip\halign{\hfil$\scriptstyle##$\hfil\cr
<\cr\approx\cr}}}
{\vcenter{\offinterlineskip\halign{\hfil$\scriptscriptstyle##$\hfil\cr
<\cr\approx\cr}}}}}

\def\goa{\mathrel{\mathchoice {\vcenter{\offinterlineskip\halign{\hfil
$\displaystyle##$\hfil\cr>\cr\approx\cr}}}
{\vcenter{\offinterlineskip\halign{\hfil$\textstyle##$\hfil\cr
>\cr\approx\cr}}}
{\vcenter{\offinterlineskip\halign{\hfil$\scriptstyle##$\hfil\cr
>\cr\approx\cr}}}
{\vcenter{\offinterlineskip\halign{\hfil$\scriptscriptstyle##$\hfil\cr
>\cr\approx\cr}}}}}

\def\diameter{{\ifmmode\mathchoice
{\ooalign{\hfil\hbox{$\displaystyle/$}\hfil\crcr
{\hbox{$\displaystyle\mathchar"20D$}}}}
{\ooalign{\hfil\hbox{$\textstyle/$}\hfil\crcr
{\hbox{$\textstyle\mathchar"20D$}}}}
{\ooalign{\hfil\hbox{$\scriptstyle/$}\hfil\crcr
{\hbox{$\scriptstyle\mathchar"20D$}}}}
{\ooalign{\hfil\hbox{$\scriptscriptstyle/$}\hfil\crcr
{\hbox{$\scriptscriptstyle\mathchar"20D$}}}}
\else{\ooalign{\hfil/\hfil\crcr\mathhexbox20D}}%
\fi}}

\def\sq{\ifmmode\squareforqed\else{\unskip\nobreak\hfil
\penalty50\hskip1em\null\nobreak\hfil\squareforqed
\parfillskip=0pt\finalhyphendemerits=0\endgraf}\fi}
\def\squareforqed{\hbox{\rlap{$\sqcap$}$\sqcup$}}


\def\bbbc{{\mathchoice {\setbox0=\hbox{$\displaystyle\rm C$}\hbox{\hbox
to0pt{\kern0.4\wd0\vrule height0.9\ht0\hss}\box0}}
{\setbox0=\hbox{$\textstyle\rm C$}\hbox{\hbox
to0pt{\kern0.4\wd0\vrule height0.9\ht0\hss}\box0}}
{\setbox0=\hbox{$\scriptstyle\rm C$}\hbox{\hbox
to0pt{\kern0.4\wd0\vrule height0.9\ht0\hss}\box0}}
{\setbox0=\hbox{$\scriptscriptstyle\rm C$}\hbox{\hbox
to0pt{\kern0.4\wd0\vrule height0.9\ht0\hss}\box0}}}}
\def\bbbq{{\mathchoice {\setbox0=\hbox{$\displaystyle\rm
Q$}\hbox{\raise
0.15\ht0\hbox to0pt{\kern0.4\wd0\vrule height0.8\ht0\hss}\box0}}
{\setbox0=\hbox{$\textstyle\rm Q$}\hbox{\raise
0.15\ht0\hbox to0pt{\kern0.4\wd0\vrule height0.8\ht0\hss}\box0}}
{\setbox0=\hbox{$\scriptstyle\rm Q$}\hbox{\raise
0.15\ht0\hbox to0pt{\kern0.4\wd0\vrule height0.7\ht0\hss}\box0}}
{\setbox0=\hbox{$\scriptscriptstyle\rm Q$}\hbox{\raise
0.15\ht0\hbox to0pt{\kern0.4\wd0\vrule height0.7\ht0\hss}\box0}}}}
\def\bbbt{{\mathchoice {\setbox0=\hbox{$\displaystyle\rm
T$}\hbox{\hbox to0pt{\kern0.3\wd0\vrule height0.9\ht0\hss}\box0}}
{\setbox0=\hbox{$\textstyle\rm T$}\hbox{\hbox
to0pt{\kern0.3\wd0\vrule height0.9\ht0\hss}\box0}}
{\setbox0=\hbox{$\scriptstyle\rm T$}\hbox{\hbox
to0pt{\kern0.3\wd0\vrule height0.9\ht0\hss}\box0}}
{\setbox0=\hbox{$\scriptscriptstyle\rm T$}\hbox{\hbox
to0pt{\kern0.3\wd0\vrule height0.9\ht0\hss}\box0}}}}
\def\bbbs{{\mathchoice
{\setbox0=\hbox{$\displaystyle     \rm S$}\hbox{\raise0.5\ht0\hbox
to0pt{\kern0.35\wd0\vrule height0.45\ht0\hss}\hbox
to0pt{\kern0.55\wd0\vrule height0.5\ht0\hss}\box0}}
{\setbox0=\hbox{$\textstyle        \rm S$}\hbox{\raise0.5\ht0\hbox
to0pt{\kern0.35\wd0\vrule height0.45\ht0\hss}\hbox
to0pt{\kern0.55\wd0\vrule height0.5\ht0\hss}\box0}}
{\setbox0=\hbox{$\scriptstyle      \rm S$}\hbox{\raise0.5\ht0\hbox
to0pt{\kern0.35\wd0\vrule height0.45\ht0\hss}\raise0.05\ht0\hbox
to0pt{\kern0.5\wd0\vrule height0.45\ht0\hss}\box0}}
{\setbox0=\hbox{$\scriptscriptstyle\rm S$}\hbox{\raise0.5\ht0\hbox
to0pt{\kern0.4\wd0\vrule height0.45\ht0\hss}\raise0.05\ht0\hbox
to0pt{\kern0.55\wd0\vrule height0.45\ht0\hss}\box0}}}}
\def\bbbz{{\mathchoice {\hbox{$\sf\textstyle Z\kern-0.4em Z$}}
{\hbox{$\sf\textstyle Z\kern-0.4em Z$}}
{\hbox{$\sf\scriptstyle Z\kern-0.3em Z$}}
{\hbox{$\sf\scriptscriptstyle Z\kern-0.2em Z$}}}}


\ifprod@font
  \mathchardef\la="3\@xm2E
  \mathchardef\getsto="3\@xm1C
  \mathchardef\lid="3\@xm35
  \mathchardef\grole="3\@xm3F
  \mathchardef\loa="3\@xm2F
  \mathchardef\ga="3\@xm26
  \mathchardef\gid="3\@xm3D
  \mathchardef\leogr="3\@xm37
  \mathchardef\goa="3\@xm27
  \mathchardef\sq="0\@xm03
%
%
\def\diameter{{%
  \ifmmode
    \mathchoice
    {\ooalign{\hfil\hbox{$\displaystyle/$}\hfil\crcr
    {\lower.2ex\hbox{$\displaystyle\mathchar"20D$}}}}%
    {\ooalign{\hfil\hbox{$\textstyle/$}\hfil\crcr
    {\lower.2ex\hbox{$\textstyle\mathchar"20D$}}}}%
    {\ooalign{\hfil\hbox{$\scriptstyle/$}\hfil\crcr
    {\lower.1ex\hbox{$\scriptstyle\mathchar"20D$}}}}%
    {\ooalign{\hfil\hbox{$\scriptscriptstyle/$}\hfil\crcr
    {\lower.1ex\hbox{$\scriptscriptstyle\mathchar"20D$}}}}%
  \else
    {\ooalign{\hfil/\hfil\crcr\lower.2ex\hbox{\mathhexbox20D}}}%
  \fi
}}
%
%

\def\bbbc{{\Bbb{C}}}
\def\bbbq{{\Bbb{Q}}}
\def\bbbt{{\Bbb{T}}}
\def\bbbs{{\Bbb{S}}}
\def\bbbz{{\Bbb{Z}}}
\fi


\ifprod@font
\mathchardef\boxdot="2\@xm00
\mathchardef\boxplus="2\@xm01
\mathchardef\boxtimes="2\@xm02
\mathchardef\square="0\@xm03
\mathchardef\blacksquare="0\@xm04
\mathchardef\centerdot="2\@xm05
\mathchardef\lozenge="0\@xm06
\mathchardef\blacklozenge="0\@xm07
\mathchardef\circlearrowright="3\@xm08
\mathchardef\circlearrowleft="3\@xm09
\mathchardef\rightleftharpoons="3\@xm0A
\mathchardef\leftrightharpoons="3\@xm0B
\mathchardef\boxminus="2\@xm0C
\mathchardef\Vdash="3\@xm0D
\mathchardef\Vvdash="3\@xm0E
\mathchardef\vDash="3\@xm0F
\mathchardef\twoheadrightarrow="3\@xm10
\mathchardef\twoheadleftarrow="3\@xm11
\mathchardef\leftleftarrows="3\@xm12
\mathchardef\rightrightarrows="3\@xm13
\mathchardef\upuparrows="3\@xm14
\mathchardef\downdownarrows="3\@xm15
\mathchardef\upharpoonright="3\@xm16

\mathchardef\downharpoonright="3\@xm17
\mathchardef\upharpoonleft="3\@xm18
\mathchardef\downharpoonleft="3\@xm19
\mathchardef\rightarrowtail="3\@xm1A
\mathchardef\leftarrowtail="3\@xm1B
\mathchardef\leftrightarrows="3\@xm1C
\mathchardef\rightleftarrows="3\@xm1D
\mathchardef\Lsh="3\@xm1E
\mathchardef\Rsh="3\@xm1F
\mathchardef\rightsquigarrow="3\@xm20
\mathchardef\leftrightsquigarrow="3\@xm21
\mathchardef\looparrowleft="3\@xm22
\mathchardef\looparrowright="3\@xm23
\mathchardef\circeq="3\@xm24
\mathchardef\succsim="3\@xm25
\mathchardef\gtrsim="3\@xm26
\mathchardef\gtrapprox="3\@xm27
\mathchardef\multimap="3\@xm28
\mathchardef\therefore="3\@xm29
\mathchardef\because="3\@xm2A
\mathchardef\doteqdot="3\@xm2B

\mathchardef\triangleq="3\@xm2C
\mathchardef\precsim="3\@xm2D
\mathchardef\lesssim="3\@xm2E
\mathchardef\lessapprox="3\@xm2F
\mathchardef\eqslantless="3\@xm30
\mathchardef\eqslantgtr="3\@xm31
\mathchardef\curlyeqprec="3\@xm32
\mathchardef\curlyeqsucc="3\@xm33
\mathchardef\preccurlyeq="3\@xm34
\mathchardef\leqq="3\@xm35
\mathchardef\leqslant="3\@xm36
\mathchardef\lessgtr="3\@xm37
\mathchardef\backprime="0\@xm38
\mathchardef\risingdotseq="3\@xm3A
\mathchardef\fallingdotseq="3\@xm3B
\mathchardef\succcurlyeq="3\@xm3C
\mathchardef\geqq="3\@xm3D
\mathchardef\geqslant="3\@xm3E
\mathchardef\gtrless="3\@xm3F
\mathchardef\sqsubset="3\@xm40
\mathchardef\sqsupset="3\@xm41
\mathchardef\vartriangleright="3\@xm42
\mathchardef\vartriangleleft="3\@xm43
\mathchardef\trianglerighteq="3\@xm44
\mathchardef\trianglelefteq="3\@xm45
\mathchardef\bigstar="0\@xm46
\mathchardef\between="3\@xm47
\mathchardef\blacktriangledown="0\@xm48
\mathchardef\blacktriangleright="3\@xm49
\mathchardef\blacktriangleleft="3\@xm4A
\mathchardef\vartriangle="0\@xm4D
\mathchardef\blacktriangle="0\@xm4E
\mathchardef\triangledown="0\@xm4F
\mathchardef\eqcirc="3\@xm50
\mathchardef\lesseqgtr="3\@xm51
\mathchardef\gtreqless="3\@xm52
\mathchardef\lesseqqgtr="3\@xm53
\mathchardef\gtreqqless="3\@xm54
\mathchardef\Rrightarrow="3\@xm56
\mathchardef\Lleftarrow="3\@xm57
\mathchardef\veebar="2\@xm59
\mathchardef\barwedge="2\@xm5A
\mathchardef\doublebarwedge="2\@xm5B
\mathchardef\angle="0\@xm5C
\mathchardef\measuredangle="0\@xm5D
\mathchardef\sphericalangle="0\@xm5E
\mathchardef\varpropto="3\@xm5F
\mathchardef\smallsmile="3\@xm60
\mathchardef\smallfrown="3\@xm61
\mathchardef\Subset="3\@xm62
\mathchardef\Supset="3\@xm63
\mathchardef\Cup="2\@xm64

\mathchardef\Cap="2\@xm65

\mathchardef\curlywedge="2\@xm66
\mathchardef\curlyvee="2\@xm67
\mathchardef\leftthreetimes="2\@xm68
\mathchardef\rightthreetimes="2\@xm69
\mathchardef\subseteqq="3\@xm6A
\mathchardef\supseteqq="3\@xm6B
\mathchardef\bumpeq="3\@xm6C
\mathchardef\Bumpeq="3\@xm6D
\mathchardef\lll="3\@xm6E

\mathchardef\ggg="3\@xm6F

\mathchardef\circledS="0\@xm73
\mathchardef\pitchfork="3\@xm74
\mathchardef\dotplus="2\@xm75
\mathchardef\backsim="3\@xm76
\mathchardef\backsimeq="3\@xm77
\mathchardef\complement="0\@xm7B
\mathchardef\intercal="2\@xm7C
\mathchardef\circledcirc="2\@xm7D
\mathchardef\circledast="2\@xm7E
\mathchardef\circleddash="2\@xm7F
\def\ulcorner{\delimiter"4\@xm70\@xm70 }
\def\urcorner{\delimiter"5\@xm71\@xm71 }
\def\llcorner{\delimiter"4\@xm78\@xm78 }
\def\lrcorner{\delimiter"5\@xm79\@xm79 }
\def\yen{\mathhexbox\@xm55 }
\def\checkmark{\mathhexbox\@xm58 }
\def\circledR{\mathhexbox\@xm72 }
\def\maltese{\mathhexbox\@xm7A }
\mathchardef\lvertneqq="3\@ym00
\mathchardef\gvertneqq="3\@ym01
\mathchardef\nleq="3\@ym02
\mathchardef\ngeq="3\@ym03
\mathchardef\nless="3\@ym04
\mathchardef\ngtr="3\@ym05
\mathchardef\nprec="3\@ym06
\mathchardef\nsucc="3\@ym07
\mathchardef\lneqq="3\@ym08
\mathchardef\gneqq="3\@ym09
\mathchardef\nleqslant="3\@ym0A
\mathchardef\ngeqslant="3\@ym0B
\mathchardef\lneq="3\@ym0C
\mathchardef\gneq="3\@ym0D
\mathchardef\npreceq="3\@ym0E
\mathchardef\nsucceq="3\@ym0F
\mathchardef\precnsim="3\@ym10
\mathchardef\succnsim="3\@ym11
\mathchardef\lnsim="3\@ym12
\mathchardef\gnsim="3\@ym13
\mathchardef\nleqq="3\@ym14
\mathchardef\ngeqq="3\@ym15
\mathchardef\precneqq="3\@ym16
\mathchardef\succneqq="3\@ym17
\mathchardef\precnapprox="3\@ym18
\mathchardef\succnapprox="3\@ym19
\mathchardef\lnapprox="3\@ym1A
\mathchardef\gnapprox="3\@ym1B
\mathchardef\nsim="3\@ym1C
\mathchardef\ncong="3\@ym1D

\mathchardef\varsubsetneq="3\@ym20
\mathchardef\varsupsetneq="3\@ym21
\mathchardef\nsubseteqq="3\@ym22
\mathchardef\nsupseteqq="3\@ym23
\mathchardef\subsetneqq="3\@ym24
\mathchardef\supsetneqq="3\@ym25
\mathchardef\varsubsetneqq="3\@ym26
\mathchardef\varsupsetneqq="3\@ym27
\mathchardef\subsetneq="3\@ym28
\mathchardef\supsetneq="3\@ym29
\mathchardef\nsubseteq="3\@ym2A
\mathchardef\nsupseteq="3\@ym2B
\mathchardef\nparallel="3\@ym2C
\mathchardef\nmid="3\@ym2D
\mathchardef\nshortmid="3\@ym2E
\mathchardef\nshortparallel="3\@ym2F
\mathchardef\nvdash="3\@ym30
\mathchardef\nVdash="3\@ym31
\mathchardef\nvDash="3\@ym32
\mathchardef\nVDash="3\@ym33
\mathchardef\ntrianglerighteq="3\@ym34
\mathchardef\ntrianglelefteq="3\@ym35
\mathchardef\ntriangleleft="3\@ym36
\mathchardef\ntriangleright="3\@ym37
\mathchardef\nleftarrow="3\@ym38
\mathchardef\nrightarrow="3\@ym39
\mathchardef\nLeftarrow="3\@ym3A
\mathchardef\nRightarrow="3\@ym3B
\mathchardef\nLeftrightarrow="3\@ym3C
\mathchardef\nleftrightarrow="3\@ym3D
\mathchardef\divideontimes="2\@ym3E
\mathchardef\varnothing="0\@ym3F
\mathchardef\nexists="0\@ym40
\mathchardef\mho="0\@ym66
\mathchardef\eth="0\@ym67
\mathchardef\eqsim="3\@ym68
\mathchardef\beth="0\@ym69
\mathchardef\gimel="0\@ym6A
\mathchardef\daleth="0\@ym6B
\mathchardef\lessdot="3\@ym6C
\mathchardef\gtrdot="3\@ym6D
\mathchardef\ltimes="2\@ym6E
\mathchardef\rtimes="2\@ym6F
\mathchardef\shortmid="3\@ym70
\mathchardef\shortparallel="3\@ym71
\mathchardef\smallsetminus="2\@ym72
\mathchardef\thicksim="3\@ym73
\mathchardef\thickapprox="3\@ym74
\mathchardef\approxeq="3\@ym75
\mathchardef\succapprox="3\@ym76
\mathchardef\precapprox="3\@ym77
\mathchardef\curvearrowleft="3\@ym78
\mathchardef\curvearrowright="3\@ym79
\mathchardef\digamma="0\@ym7A
\mathchardef\varkappa="0\@ym7B
\mathchardef\hslash="0\@ym7D
\mathchardef\hbar="0\@ym7E
\mathchardef\backepsilon="3\@ym7F


\def\Bbb{\ifmmode\let\next\Bbb@\else
\def\next{\errmessage{Use \string\Bbb\space only in math mode}}\fi\next}
\def\Bbb@#1{{\Bbb@@{#1}}}
\def\Bbb@@#1{\fam\ymfam#1}
\fi


\def\Nulle{0} 
\def\Afe{1}   
\def\Hae{2}   
\def\Hbe{3}   
\def\Hce{4}   
\def\Hde{5}   


\newcount\LastMac       \LastMac=\Nulle

\newskip\half      \half=5.5pt plus 1.5pt minus 2.25pt
\newskip\one       \one=11pt plus 3pt minus 5.5pt
\newskip\onehalf   \onehalf=16.5pt plus 5.5pt minus 8.25pt
\newskip\two       \two=22pt plus 5.5pt minus 11pt

\def\Half{\addvspace{\half}}
\def\One{\addvspace{\one}}
\def\OneHalf{\addvspace{\onehalf}}
\def\Two{\addvspace{\two}}


\def\Raggedright{
  \rightskip=\z@ plus \hsize\relax
}

\def\Fullout{
  \rightskip=\z@\relax
}

\def\Hang#1#2{
  \hangindent=#1%
  \hangafter=#2\relax
}


\newif\ifsp@page
\def\pagestyle#1{\csname ps@#1\endcsname}
\def\thispagestyle#1{\global\sp@pagetrue\gdef\sp@type{#1}}

\def\ps@titlepage{%
  \def\@oddhead{\eightpoint\noindent \the\CatchLine
    \ifprod@font\else\qquad Printed\ \today\fi \hfil}%
  \let\@evenhead=\@oddhead
}

\def\ps@headings{%
  \def\@oddhead{\elevenpoint\it\noindent
    \hfill\the\RightHeader\hskip1.5em\rm\folio}%
  \def\@evenhead{\elevenpoint\noindent
    \folio\hskip1.5em\it\the\LeftHeader\hfill}%
}

\def\ps@plate{%
  \def\@oddhead{\eightpoint\noindent\plt@cap\hfil}%
  \def\@evenhead{\eightpoint\noindent\plt@cap\hfil}%
}



\def\title#1{
  \bgroup
    \vbox to 8pt{\vss}%
    \seventeenpoint
    \Raggedright
    \noindent \strut{\bf #1}\par
  \egroup
}

\def\author#1{
  \bgroup
    \ifnum\LastMac=\Afe \OneHalf\else \vskip 21pt\fi
    \fourteenpoint
    \Raggedright
    \noindent \strut #1\par
    \vskip 3pt%
  \egroup
}

\def\affiliation#1{
  \bgroup
    \vskip -4pt%
    \eightpoint
    \Raggedright
    \noindent \strut {\it #1}\par
  \egroup
  \LastMac=\Afe\relax
}

\def\acceptedline#1{
  \bgroup
    \Two
    \eightpoint
    \Raggedright
    \noindent \strut #1\par
  \egroup
}

\long\def\abstract#1{%
  \bgroup
    \vskip 20pt%
    \everypar{\Hang{11pc}{0}}%
    \noindent{\ninebf ABSTRACT}\par
    \tenpoint
    \Fullout
    \noindent #1\par
  \egroup
}

\long\def\keywords#1{
  \bgroup
    \Half
    \everypar{\Hang{11pc}{0}}%
    \tenpoint
    \Fullout
    \noindent\hbox{\bf Key words:}\ #1\par
  \egroup
}


\def\maketitle{%
  \EndOpening
  \ifsinglecol \else \MakePage\fi
}



\def\Autonumber{
  \global\AutoNumbertrue  
}

\newif\ifAutoNumber \AutoNumberfalse
\newcount\Sec        
\newcount\SecSec
\newcount\SecSecSec

\Sec=\z@

\def\:{\let\@sptoken= } \:  
\def\:{\@xifnch} \expandafter\def\: {\futurelet\@tempc\@ifnch}

\def\@ifnextchar#1#2#3{%
  \let\@tempMACe #1%
  \def\@tempMACa{#2}%
  \def\@tempMACb{#3}%
  \futurelet \@tempMACc\@ifnch%
}

\def\@ifnch{%
\ifx \@tempMACc \@sptoken%
  \let\@tempMACd\@xifnch%
\else%
  \ifx \@tempMACc \@tempMACe%
    \let\@tempMACd\@tempMACa%
  \else%
    \let\@tempMACd\@tempMACb%
  \fi%
\fi%
\@tempMACd%
}

\def\@ifstar#1#2{\@ifnextchar *{\def\@tempMACa*{#1}\@tempMACa}{#2}}

\newskip\@tempskipb

\def\addvspace#1{%
  \ifvmode\else \endgraf\fi%
  \ifdim\lastskip=\z@%
    \vskip #1\relax%
  \else%
    \@tempskipb#1\relax\@xaddvskip%
  \fi%
}

\def\@xaddvskip{%
  \ifdim\lastskip<\@tempskipb%
    \vskip-\lastskip%
    \vskip\@tempskipb\relax%
  \else%
    \ifdim\@tempskipb<\z@%
      \ifdim\lastskip<\z@ \else%
        \advance\@tempskipb\lastskip%
        \vskip-\lastskip\vskip\@tempskipb%
      \fi%
    \fi%
  \fi%
}

\newskip\@tmpSKIP

\def\addpen#1{%
  \ifvmode
    \if@nobreak
    \else
      \ifdim\lastskip=\z@
        \penalty#1\relax
      \else
        \@tmpSKIP=\lastskip
        \vskip -\lastskip
        \penalty#1\vskip\@tmpSKIP
      \fi
    \fi
  \fi
}

\newcount\@clubpen   \@clubpen=\clubpenalty
\newif\if@nobreak    \@nobreakfalse

\def\@noafterindent{%
  \global\@nobreaktrue
  \everypar{\if@nobreak
              \global\@nobreakfalse
              \clubpenalty \@M
              {\setbox\z@\lastbox}%
              \LastMac=\Nulle\relax%
            \else
              \clubpenalty \@clubpen
              \everypar{}%
            \fi}
}

\newcount\gds@cbrk   \gds@cbrk=-300

\def\@nohdbrk{\interlinepenalty \@M\relax}

\let\@par=\par
\def\@restorepar{\def\par{\@par}}

\newif\if@endpe   \@endpefalse
 
\def\@doendpe{\@endpetrue \@nobreakfalse \LastMac=\Nulle\relax%
     \def\par{\@restorepar\everypar{}\par\@endpefalse}%
              \everypar{\setbox\z@\lastbox\everypar{}\@endpefalse}%
}

\def\section{\@ifstar{\@ssection}{\@section}}

\def\@section#1{
  \if@nobreak
    \everypar{}%
    \ifnum\LastMac=\Hae \addvspace{\half}\fi
  \else
    \addpen{\gds@cbrk}%
    \addvspace{\two}%
  \fi
  \bgroup
    \ninepoint\bf
    \Raggedright
    \ifAutoNumber
      \global\advance\Sec \@ne
      \noindent\@nohdbrk\number\Sec\hskip 1pc \uppercase{#1}\par
      \global\SecSec=\z@
    \else
      \noindent\@nohdbrk\uppercase{#1}\par
    \fi
  \egroup
  \nobreak
  \vskip\half
  \nobreak
  \@noafterindent
  \LastMac=\Hae\relax
}

\def\@ssection#1{
  \if@nobreak
    \everypar{}%
    \ifnum\LastMac=\Hae \addvspace{\half}\fi
  \else
    \addpen{\gds@cbrk}%
    \addvspace{\two}%
  \fi
  \bgroup
    \ninepoint\bf
    \Raggedright
    \noindent\@nohdbrk\uppercase{#1}\par
  \egroup
  \nobreak
  \vskip\half
  \nobreak
  \@noafterindent
  \LastMac=\Hae\relax
}

\def\subsection#1{
  \if@nobreak
    \everypar{}%
    \ifnum\LastMac=\Hae \addvspace{1pt plus 1pt minus .5pt}\fi
  \else
    \addpen{\gds@cbrk}%
    \addvspace{\onehalf}%
  \fi
  \bgroup
    \ninepoint\bf
    \Raggedright
    \ifAutoNumber
      \global\advance\SecSec \@ne
      \noindent\@nohdbrk\number\Sec.\number\SecSec \hskip 1pc\relax #1\par
      \global\SecSecSec=\z@
    \else
      \noindent\@nohdbrk #1\par
    \fi
  \egroup
  \nobreak
  \vskip\half
  \nobreak
  \@noafterindent
  \LastMac=\Hbe\relax
}

\def\subsubsection#1{
  \if@nobreak
    \everypar{}%
    \ifnum\LastMac=\Hbe \addvspace{1pt plus 1pt minus .5pt}\fi
  \else
    \addpen{\gds@cbrk}%
    \addvspace{\onehalf}%
  \fi
  \bgroup
    \ninepoint\it
    \Raggedright
    \ifAutoNumber
      \global\advance\SecSecSec \@ne
      \noindent\@nohdbrk\number\Sec.\number\SecSec.\number\SecSecSec
        \hskip 1pc\relax #1\par
    \else
      \noindent\@nohdbrk #1\par
    \fi
  \egroup
  \nobreak
  \vskip\half
  \nobreak
  \@noafterindent
  \LastMac=\Hce\relax
}

\def\paragraph#1{
  \if@nobreak
    \everypar{}%
  \else
    \addpen{\gds@cbrk}%
    \addvspace{\one}%
  \fi%
  \bgroup%
    \ninepoint\it
    \noindent #1\ \nobreak%
  \egroup
  \LastMac=\Hde\relax
  \ignorespaces
}




\def\beginlist{%
  \par\if@nobreak \else\addvspace{\half}\fi%
  \bgroup%
    \ninepoint
    \let\item=\list@item%
}

\def\list@item{%
  \par\noindent\hskip 1em\relax%
  \ignorespaces%
}

\def\endlist{\par\egroup\addvspace{\half}\@doendpe}


\def\beginrefs{%
  \par
  \bgroup
    \eightpoint
    \Raggedright
    \let\bibitem=\bib@item
}

\def\bib@item{%
  \par\parindent=1.5em\Hang{1.5em}{1}%
  \everypar={\Hang{1.5em}{1}\ignorespaces}%
  \noindent\ignorespaces
}

\def\endrefs{\par\egroup\@doendpe}


\newtoks\CatchLine

\def\@journal{Mon.\ Not.\ R.\ Astron.\ Soc.\ }  
\def\@pubyear{1994}        
\def\@pagerange{000--000}  
\def\@volume{000}          
\def\@microfiche{}         %

\def\pubyear#1{\gdef\@pubyear{#1}\@makecatchline}
\def\pagerange#1{\gdef\@pagerange{#1}\@makecatchline}
\def\volume#1{\gdef\@volume{#1}\@makecatchline}
\def\microfiche#1{\gdef\@microfiche{and Microfiche\ #1}\@makecatchline}

\def\@makecatchline{%
  \global\CatchLine{%
    {\rm \@journal {\bf \@volume},\ \@pagerange\ (\@pubyear)\ \@microfiche}}%
}

\@makecatchline 

\newtoks\LeftHeader
\def\shortauthor#1{
  \global\LeftHeader{#1}%
}

\newtoks\RightHeader
\def\shorttitle#1{
  \global\RightHeader{#1}%
}

\def\PageHead{
  \begingroup
    \ifsp@page
      \csname ps@\sp@type\endcsname
      \global\sp@pagefalse
    \fi
    \ifodd\pageno
      \let\the@head=\@oddhead
    \else
      \let\the@head=\@evenhead
    \fi
    \vbox to \z@{\vskip-22.5\p@%
      \hbox to \PageWidth{\vbox to8.5\p@{}%
        \the@head
      }%
    \vss}%
  \endgroup
  \nointerlineskip
}

\def\today{%
  \number\day\space
  \ifcase\month\or January\or February\or March\or April\or May\or June\or
    July\or August\or September\or October\or November\or December\fi
  \space\number\year%
}

\def\PageFoot{} 

\def\authorcomment#1{%
  \gdef\PageFoot{%
    \nointerlineskip%
    \vbox to 22pt{\vfil%
      \hbox to \PageWidth{\elevenpoint\noindent \hfil #1 \hfil}}%
  }%
}


\newif\ifplate@page
\newbox\plt@box

\def\beginplatepage{%
  \let\plate=\plate@head
  \let\caption=\fig@caption
  \global\setbox\plt@box=\vbox\bgroup
  \TEMPDIMEN=\PageWidth 
  \hsize=\PageWidth\relax
}

\def\endplatepage{\par\egroup\global\plate@pagetrue}
\def\plate@head#1{\gdef\plt@cap{#1}}


\def\letters{%
  \gdef\folio{\ifnum\pageno<\z@ L\romannumeral-\pageno
    \else L\number\pageno \fi}%
}


\everydisplay{\displaysetup}

\newif\ifeqno
\newif\ifleqno

\def\displaysetup#1$${%
 \displaytest#1\eqno\eqno\displaytest
}

\def\displaytest#1\eqno#2\eqno#3\displaytest{%
 \if!#3!\ldisplaytest#1\leqno\leqno\ldisplaytest
 \else\eqnotrue\leqnofalse\def\eqn{#2}\def\eq{#1}\fi
 \generaldisplay$$}

\def\ldisplaytest#1\leqno#2\leqno#3\ldisplaytest{%
 \def\eq{#1}%
 \if!#3!\eqnofalse\else\eqnotrue\leqnotrue
  \def\eqn{#2}\fi}

\def\generaldisplay{%
\ifeqno \ifleqno 
   \hbox to \hsize{\noindent
     $\displaystyle\eq$\hfil$\displaystyle\eqn$}
  \else
    \hbox to \hsize{\noindent
     $\displaystyle\eq$\hfil$\displaystyle\eqn$}
  \fi
 \else
 \hbox to \hsize{\vbox{\noindent
  $\displaystyle\eq$\hfil}}
 \fi
}


\def\@notice{%
  \par\Two%
  \noindent{\b@ls{11pt}\ninerm This paper has been produced using the
    Blackwell Scientific Publications \TeX\ macros.\par}%
}

\outer\def\bye{\@notice\par\vfill\supereject\end}


\def\start@mess{%
  Monthly notices of the RAS journal style (\@typeface)\space
    v\@version,\space \@verdate.%
}

\everyjob{\Warn{\start@mess}}



\newif\if@debug \@debugfalse  

\def\Print#1{\if@debug\immediate\write16{#1}\else \fi}
\def\Warn#1{\immediate\write16{#1}}
\def\wlog#1{}

\newcount\Iteration 

\def\Single{0} \def\Double{1}                 
\def\Figure{0} \def\Table{1}                  

\def\InStack{0}  
\def\InZoneA{1}
\def\InZoneB{2}
\def\InZoneC{3}

\newcount\TEMPCOUNT 
\newdimen\TEMPDIMEN 
\newbox\TEMPBOX     
\newbox\VOIDBOX     

\newcount\LengthOfStack 
\newcount\MaxItems      
\newcount\StackPointer
\newcount\Point         
\newcount\NextFigure    
\newcount\NextTable     
\newcount\NextItem      

\newcount\StatusStack   
\newcount\NumStack      
\newcount\TypeStack     
\newcount\SpanStack     
\newcount\BoxStack      

\newcount\ItemSTATUS    
\newcount\ItemNUMBER    
\newcount\ItemTYPE      
\newcount\ItemSPAN      
\newbox\ItemBOX         
\newdimen\ItemSIZE      

\newdimen\PageHeight    
\newdimen\TextLeading   
\newdimen\Feathering    
\newcount\LinesPerPage  
\newdimen\ColumnWidth   
\newdimen\ColumnGap     
\newdimen\PageWidth     
\newdimen\BodgeHeight   
\newcount\Leading       

\newdimen\ZoneBSize  
\newdimen\TextSize   
\newbox\ZoneABOX     
\newbox\ZoneBBOX     
\newbox\ZoneCBOX     

\newif\ifFirstSingleItem
\newif\ifFirstZoneA
\newif\ifMakePageInComplete
\newif\ifMoreFigures \MoreFiguresfalse 
\newif\ifMoreTables  \MoreTablesfalse  

\newif\ifFigInZoneB 
\newif\ifFigInZoneC 
\newif\ifTabInZoneB 
\newif\ifTabInZoneC

\newif\ifZoneAFullPage

\newbox\MidBOX    
\newbox\LeftBOX
\newbox\RightBOX
\newbox\PageBOX   

\newif\ifLeftCOL  
\LeftCOLtrue

\newdimen\ZoneBAdjust

\newcount\ItemFits
\def\Yes{1}
\def\No{2}


\MaxItems=15
\NextFigure=\z@        
\NextTable=\@ne

\BodgeHeight=6pt
\TextLeading=11pt    
\Leading=11
\Feathering=\z@      
\LinesPerPage=61     
\topskip=\TextLeading
\ColumnWidth=20pc    
\ColumnGap=2pc       

\newskip\ItemSepamount  
\ItemSepamount=\TextLeading plus \TextLeading minus 4pt

\parskip=\z@ plus .1pt
\parindent=18pt
\widowpenalty=\z@
\clubpenalty=10000
\tolerance=1500
\hbadness=1500
\abovedisplayskip=6pt plus 2pt minus 2pt
\belowdisplayskip=6pt plus 2pt minus 2pt
\abovedisplayshortskip=6pt plus 2pt minus 2pt
\belowdisplayshortskip=6pt plus 2pt minus 2pt

\ninepoint 


\PageHeight=682pt

\PageWidth=2\ColumnWidth
\advance\PageWidth by \ColumnGap

\pagestyle{headings}




\newcount\DUMMY \StatusStack=\allocationnumber
\newcount\DUMMY \newcount\DUMMY \newcount\DUMMY 
\newcount\DUMMY \newcount\DUMMY \newcount\DUMMY 
\newcount\DUMMY \newcount\DUMMY \newcount\DUMMY
\newcount\DUMMY \newcount\DUMMY \newcount\DUMMY 
\newcount\DUMMY \newcount\DUMMY \newcount\DUMMY

\newcount\DUMMY \NumStack=\allocationnumber
\newcount\DUMMY \newcount\DUMMY \newcount\DUMMY 
\newcount\DUMMY \newcount\DUMMY \newcount\DUMMY 
\newcount\DUMMY \newcount\DUMMY \newcount\DUMMY 
\newcount\DUMMY \newcount\DUMMY \newcount\DUMMY 
\newcount\DUMMY \newcount\DUMMY \newcount\DUMMY

\newcount\DUMMY \TypeStack=\allocationnumber
\newcount\DUMMY \newcount\DUMMY \newcount\DUMMY 
\newcount\DUMMY \newcount\DUMMY \newcount\DUMMY 
\newcount\DUMMY \newcount\DUMMY \newcount\DUMMY 
\newcount\DUMMY \newcount\DUMMY \newcount\DUMMY 
\newcount\DUMMY \newcount\DUMMY \newcount\DUMMY

\newcount\DUMMY \SpanStack=\allocationnumber
\newcount\DUMMY \newcount\DUMMY \newcount\DUMMY 
\newcount\DUMMY \newcount\DUMMY \newcount\DUMMY 
\newcount\DUMMY \newcount\DUMMY \newcount\DUMMY 
\newcount\DUMMY \newcount\DUMMY \newcount\DUMMY 
\newcount\DUMMY \newcount\DUMMY \newcount\DUMMY

\newbox\DUMMY   \BoxStack=\allocationnumber
\newbox\DUMMY   \newbox\DUMMY \newbox\DUMMY 
\newbox\DUMMY   \newbox\DUMMY \newbox\DUMMY 
\newbox\DUMMY   \newbox\DUMMY \newbox\DUMMY 
\newbox\DUMMY   \newbox\DUMMY \newbox\DUMMY 
\newbox\DUMMY   \newbox\DUMMY \newbox\DUMMY

\def\wlog{\immediate\write\m@ne}


\def\GetItemAll#1{%
 \GetItemSTATUS{#1}
 \GetItemNUMBER{#1}
 \GetItemTYPE{#1}
 \GetItemSPAN{#1}
 \GetItemBOX{#1}
}

\def\GetItemSTATUS#1{%
 \Point=\StatusStack
 \advance\Point by #1
 \global\ItemSTATUS=\count\Point
}

\def\GetItemNUMBER#1{%
 \Point=\NumStack
 \advance\Point by #1
 \global\ItemNUMBER=\count\Point
}

\def\GetItemTYPE#1{%
 \Point=\TypeStack
 \advance\Point by #1
 \global\ItemTYPE=\count\Point
}

\def\GetItemSPAN#1{%
 \Point\SpanStack
 \advance\Point by #1
 \global\ItemSPAN=\count\Point
}

\def\GetItemBOX#1{%
 \Point=\BoxStack
 \advance\Point by #1
 \global\setbox\ItemBOX=\vbox{\copy\Point}
 \global\ItemSIZE=\ht\ItemBOX
 \global\advance\ItemSIZE by \dp\ItemBOX
 \TEMPCOUNT=\ItemSIZE
 \divide\TEMPCOUNT by \Leading
 \divide\TEMPCOUNT by 65536
 \advance\TEMPCOUNT \@ne
 \ItemSIZE=\TEMPCOUNT pt
 \global\multiply\ItemSIZE by \Leading
}


\def\JoinStack{%
 \ifnum\LengthOfStack=\MaxItems 
  \Warn{WARNING: Stack is full...some items will be lost!}
 \else
  \Point=\StatusStack
  \advance\Point by \LengthOfStack
  \global\count\Point=\ItemSTATUS
  \Point=\NumStack
  \advance\Point by \LengthOfStack
  \global\count\Point=\ItemNUMBER
  \Point=\TypeStack
  \advance\Point by \LengthOfStack
  \global\count\Point=\ItemTYPE
  \Point\SpanStack
  \advance\Point by \LengthOfStack
  \global\count\Point=\ItemSPAN
  \Point=\BoxStack
  \advance\Point by \LengthOfStack
  \global\setbox\Point=\vbox{\copy\ItemBOX}
  \global\advance\LengthOfStack \@ne
  \ifnum\ItemTYPE=\Figure 
   \global\MoreFigurestrue
  \else
   \global\MoreTablestrue
  \fi
 \fi
}


\def\LeaveStack#1{%
 {\Iteration=#1
 \loop
 \ifnum\Iteration<\LengthOfStack
  \advance\Iteration \@ne
  \GetItemSTATUS{\Iteration}
   \advance\Point by \m@ne
   \global\count\Point=\ItemSTATUS
  \GetItemNUMBER{\Iteration}
   \advance\Point by \m@ne
   \global\count\Point=\ItemNUMBER
  \GetItemTYPE{\Iteration}
   \advance\Point by \m@ne
   \global\count\Point=\ItemTYPE
  \GetItemSPAN{\Iteration}
   \advance\Point by \m@ne
   \global\count\Point=\ItemSPAN
  \GetItemBOX{\Iteration}
   \advance\Point by \m@ne
   \global\setbox\Point=\vbox{\copy\ItemBOX}
 \repeat}
 \global\advance\LengthOfStack by \m@ne
}


\newif\ifStackNotClean

\def\CleanStack{%
 \StackNotCleantrue
 {\Iteration=\z@
  \loop
   \ifStackNotClean
    \GetItemSTATUS{\Iteration}
    \ifnum\ItemSTATUS=\InStack
     \advance\Iteration \@ne
     \else
      \LeaveStack{\Iteration}
    \fi
   \ifnum\LengthOfStack<\Iteration
    \StackNotCleanfalse
   \fi
 \repeat}
}


\def\FindItem#1#2{%
 \global\StackPointer=\m@ne 
 {\Iteration=\z@
  \loop
  \ifnum\Iteration<\LengthOfStack
   \GetItemSTATUS{\Iteration}
   \ifnum\ItemSTATUS=\InStack
    \GetItemTYPE{\Iteration}
    \ifnum\ItemTYPE=#1
     \GetItemNUMBER{\Iteration}
     \ifnum\ItemNUMBER=#2
      \global\StackPointer=\Iteration
      \Iteration=\LengthOfStack 
     \fi
    \fi
   \fi
  \advance\Iteration \@ne
 \repeat}
}


\def\FindNext{%
 \global\StackPointer=\m@ne 
 {\Iteration=\z@
  \loop
  \ifnum\Iteration<\LengthOfStack
   \GetItemSTATUS{\Iteration}
   \ifnum\ItemSTATUS=\InStack
    \GetItemTYPE{\Iteration}
   \ifnum\ItemTYPE=\Figure
    \ifMoreFigures
      \global\NextItem=\Figure
      \global\StackPointer=\Iteration
      \Iteration=\LengthOfStack 
    \fi
   \fi
   \ifnum\ItemTYPE=\Table
    \ifMoreTables
      \global\NextItem=\Table
      \global\StackPointer=\Iteration
      \Iteration=\LengthOfStack 
    \fi
   \fi
  \fi
  \advance\Iteration \@ne
 \repeat}
}


\def\ChangeStatus#1#2{%
 \Point=\StatusStack
 \advance\Point by #1
 \global\count\Point=#2
}



\def\Zone{\InZoneA}

\ZoneBAdjust=\z@

\def\MakePage{
 \global\ZoneBSize=\PageHeight
 \global\TextSize=\ZoneBSize
 \global\ZoneAFullPagefalse
 \global\topskip=\TextLeading
 \MakePageInCompletetrue
 \MoreFigurestrue
 \MoreTablestrue
 \FigInZoneBfalse
 \FigInZoneCfalse
 \TabInZoneBfalse
 \TabInZoneCfalse
 \global\FirstSingleItemtrue
 \global\FirstZoneAtrue
 \global\setbox\ZoneABOX=\box\VOIDBOX
 \global\setbox\ZoneBBOX=\box\VOIDBOX
 \global\setbox\ZoneCBOX=\box\VOIDBOX
 \loop
  \ifMakePageInComplete
 \FindNext
 \ifnum\StackPointer=\m@ne
  \NextItem=\m@ne
  \MoreFiguresfalse
  \MoreTablesfalse
 \fi
 \ifnum\NextItem=\Figure
   \FindItem{\Figure}{\NextFigure}
   \ifnum\StackPointer=\m@ne \global\MoreFiguresfalse
   \else
    \GetItemSPAN{\StackPointer}
    \ifnum\ItemSPAN=\Single \def\Zone{\InZoneB}\relax
     \ifFigInZoneC \global\MoreFiguresfalse\fi
    \else
     \def\Zone{\InZoneA}
     \ifFigInZoneB \def\Zone{\InZoneC}\fi
    \fi
   \fi
   \ifMoreFigures\Print{}\FigureItems\fi
 \fi
\ifnum\NextItem=\Table
   \FindItem{\Table}{\NextTable}
   \ifnum\StackPointer=\m@ne \global\MoreTablesfalse
   \else
    \GetItemSPAN{\StackPointer}
    \ifnum\ItemSPAN=\Single\relax
     \ifTabInZoneC \global\MoreTablesfalse\fi
    \else
     \def\Zone{\InZoneA}
     \ifTabInZoneB \def\Zone{\InZoneC}\fi
    \fi
   \fi
   \ifMoreTables\Print{}\TableItems\fi
 \fi
   \MakePageInCompletefalse 
   \ifMoreFigures\MakePageInCompletetrue\fi
   \ifMoreTables\MakePageInCompletetrue\fi
 \repeat
 \ifZoneAFullPage
  \global\TextSize=\z@
  \global\ZoneBSize=\z@
  \global\vsize=\z@\relax
  \global\topskip=\z@\relax
  \vbox to \z@{\vss}
  \eject
 \else
 \global\advance\ZoneBSize by -\ZoneBAdjust
 \global\vsize=\ZoneBSize
 \global\hsize=\ColumnWidth
 \global\ZoneBAdjust=\z@
 \ifdim\TextSize<23pt
 \Warn{}
 \Warn{* Making column fall short: TextSize=\the\TextSize *}
 \vskip-\lastskip\eject\fi
 \fi
}

\def\MakeRightCol{
 \global\TextSize=\ZoneBSize
 \MakePageInCompletetrue
 \MoreFigurestrue
 \MoreTablestrue
 \global\FirstSingleItemtrue
 \global\setbox\ZoneBBOX=\box\VOIDBOX
 \def\Zone{\InZoneB}
 \loop
  \ifMakePageInComplete
 \FindNext
 \ifnum\StackPointer=\m@ne
  \NextItem=\m@ne
  \MoreFiguresfalse
  \MoreTablesfalse
 \fi
 \ifnum\NextItem=\Figure
   \FindItem{\Figure}{\NextFigure}
   \ifnum\StackPointer=\m@ne \MoreFiguresfalse
   \else
    \GetItemSPAN{\StackPointer}
    \ifnum\ItemSPAN=\Double\relax
     \MoreFiguresfalse\fi
   \fi
   \ifMoreFigures\Print{}\FigureItems\fi
 \fi
 \ifnum\NextItem=\Table
   \FindItem{\Table}{\NextTable}
   \ifnum\StackPointer=\m@ne \MoreTablesfalse
   \else
    \GetItemSPAN{\StackPointer}
    \ifnum\ItemSPAN=\Double\relax
     \MoreTablesfalse\fi
   \fi
   \ifMoreTables\Print{}\TableItems\fi
 \fi
   \MakePageInCompletefalse 
   \ifMoreFigures\MakePageInCompletetrue\fi
   \ifMoreTables\MakePageInCompletetrue\fi
 \repeat
 \ifZoneAFullPage
  \global\TextSize=\z@
  \global\ZoneBSize=\z@
  \global\vsize=\z@\relax
  \global\topskip=\z@\relax
  \vbox to \z@{\vss}
  \eject
 \else
 \global\vsize=\ZoneBSize
 \global\hsize=\ColumnWidth
 \ifdim\TextSize<23pt
 \Warn{}
 \Warn{* Making column fall short: TextSize=\the\TextSize *}
 \vskip-\lastskip\eject\fi
\fi
}

\def\FigureItems{
 \Print{Considering...}
 \ShowItem{\StackPointer}
 \GetItemBOX{\StackPointer} 
 \GetItemSPAN{\StackPointer}
  \CheckFitInZone 
  \ifnum\ItemFits=\Yes
   \ifnum\ItemSPAN=\Single
     \ChangeStatus{\StackPointer}{\InZoneB} 
     \global\FigInZoneBtrue
     \ifFirstSingleItem
      \hbox{}\vskip-\BodgeHeight
     \global\advance\ItemSIZE by \TextLeading
     \fi
     \unvbox\ItemBOX\ItemSep
     \global\FirstSingleItemfalse
     \global\advance\TextSize by -\ItemSIZE
     \global\advance\TextSize by -\TextLeading
   \else
    \ifFirstZoneA
     \global\advance\ItemSIZE by \TextLeading
     \global\FirstZoneAfalse\fi
    \global\advance\TextSize by -\ItemSIZE
    \global\advance\TextSize by -\TextLeading
    \global\advance\ZoneBSize by -\ItemSIZE
    \global\advance\ZoneBSize by -\TextLeading
    \ifFigInZoneB\relax
     \else
     \ifdim\TextSize<3\TextLeading
     \global\ZoneAFullPagetrue
     \fi
    \fi
    \ChangeStatus{\StackPointer}{\Zone}
    \ifnum\Zone=\InZoneC \global\FigInZoneCtrue\fi
  \fi
   \Print{TextSize=\the\TextSize}
   \Print{ZoneBSize=\the\ZoneBSize}
  \global\advance\NextFigure \@ne
   \Print{This figure has been placed.}
  \else
   \Print{No space available for this figure...holding over.}
   \Print{}
   \global\MoreFiguresfalse
  \fi
}

\def\TableItems{
 \Print{Considering...}
 \ShowItem{\StackPointer}
 \GetItemBOX{\StackPointer} 
 \GetItemSPAN{\StackPointer}
  \CheckFitInZone 
  \ifnum\ItemFits=\Yes
   \ifnum\ItemSPAN=\Single
    \ChangeStatus{\StackPointer}{\InZoneB}
     \global\TabInZoneBtrue
     \ifFirstSingleItem
      \hbox{}\vskip-\BodgeHeight
     \global\advance\ItemSIZE by \TextLeading
     \fi
     \unvbox\ItemBOX\ItemSep
     \global\FirstSingleItemfalse
     \global\advance\TextSize by -\ItemSIZE
     \global\advance\TextSize by -\TextLeading
   \else
    \ifFirstZoneA
    \global\advance\ItemSIZE by \TextLeading
    \global\FirstZoneAfalse\fi
    \global\advance\TextSize by -\ItemSIZE
    \global\advance\TextSize by -\TextLeading
    \global\advance\ZoneBSize by -\ItemSIZE
    \global\advance\ZoneBSize by -\TextLeading
    \ifFigInZoneB\relax
     \else
     \ifdim\TextSize<3\TextLeading
     \global\ZoneAFullPagetrue
     \fi
    \fi
    \ChangeStatus{\StackPointer}{\Zone}
    \ifnum\Zone=\InZoneC \global\TabInZoneCtrue\fi
   \fi
  \global\advance\NextTable \@ne
   \Print{This table has been placed.}
  \else
  \Print{No space available for this table...holding over.}
   \Print{}
   \global\MoreTablesfalse
  \fi
}


\def\CheckFitInZone{%
{\advance\TextSize by -\ItemSIZE
 \advance\TextSize by -\TextLeading
 \ifFirstSingleItem
  \advance\TextSize by \TextLeading
 \fi
 \ifnum\Zone=\InZoneA\relax
  \else \advance\TextSize by -\ZoneBAdjust
 \fi
 \ifdim\TextSize<3\TextLeading \global\ItemFits=\No
 \else \global\ItemFits=\Yes\fi}
}

\def\BeginOpening{%
  \thispagestyle{titlepage}%
  \global\setbox\ItemBOX=\vbox\bgroup%
    \hsize=\PageWidth%
    \hrule height \z@
    \ifsinglecol\vskip 6pt\fi 
}

\let\begintopmatter=\BeginOpening  

\def\EndOpening{%
  \One
  \egroup
  \ifsinglecol
    \box\ItemBOX%
    \vskip\TextLeading plus 2\TextLeading
    \@noafterindent
  \else
    \ItemNUMBER=\z@%
    \ItemTYPE=\Figure
    \ItemSPAN=\Double
    \ItemSTATUS=\InStack
    \JoinStack
  \fi
}


\newif\if@here  \@herefalse

\def\no@float{\global\@heretrue}
\let\nofloat=\relax 

\def\beginfigure{%
  \@ifstar{\global\@dfloattrue \@bfigure}{\global\@dfloatfalse \@bfigure}%
}

\def\@bfigure#1{%
  \par
  \if@dfloat
    \ItemSPAN=\Double
    \TEMPDIMEN=\PageWidth
  \else
    \ItemSPAN=\Single
    \TEMPDIMEN=\ColumnWidth
  \fi
  \ifsinglecol
    \TEMPDIMEN=\PageWidth
  \else
    \ItemSTATUS=\InStack
    \ItemNUMBER=#1%
    \ItemTYPE=\Figure
  \fi
  \bgroup
    \hsize=\TEMPDIMEN
    \global\setbox\ItemBOX=\vbox\bgroup
      \eightpoint\nostb@ls{10pt}%
      \let\caption=\fig@caption
      \ifsinglecol \let\nofloat=\no@float\fi
}

\def\fig@caption#1{%
  \vskip 5.5pt plus 6pt%
  \bgroup 
    \eightpoint\nostb@ls{10pt}%
    \setbox\TEMPBOX=\hbox{#1}%
    \ifdim\wd\TEMPBOX>\TEMPDIMEN
      \noindent \unhbox\TEMPBOX\par
    \else
      \hbox to \hsize{\hfil\unhbox\TEMPBOX\hfil}%
    \fi
  \egroup
}

\def\endfigure{%
  \par\egroup 
  \egroup
  \ifsinglecol
    \if@here \midinsert\global\@herefalse\else \topinsert\fi
      \unvbox\ItemBOX
    \endinsert
  \else
    \JoinStack
    \Print{Processing source for figure \the\ItemNUMBER}%
  \fi
}


\newbox\tab@cap@box
\def\tab@caption#1{\global\setbox\tab@cap@box=\hbox{#1\par}}

\newtoks\tab@txt@toks
\long\def\tab@txt#1{\global\tab@txt@toks={#1}\global\table@txttrue}

\newif\iftable@txt  \table@txtfalse
\newif\if@dfloat    \@dfloatfalse

\def\begintable{%
  \@ifstar{\global\@dfloattrue \@btable}{\global\@dfloatfalse \@btable}%
}

\def\@btable#1{%
  \par
  \if@dfloat
    \ItemSPAN=\Double
    \TEMPDIMEN=\PageWidth
  \else
    \ItemSPAN=\Single
    \TEMPDIMEN=\ColumnWidth
  \fi
  \ifsinglecol
    \TEMPDIMEN=\PageWidth
  \else
    \ItemSTATUS=\InStack
    \ItemNUMBER=#1%
    \ItemTYPE=\Table
  \fi
  \bgroup
    \eightpoint\nostb@ls{10pt}%
    \global\setbox\ItemBOX=\vbox\bgroup
      \let\caption=\tab@caption
      \let\tabletext=\tab@txt
      \ifsinglecol \let\nofloat=\no@float\fi
}

\def\endtable{%
  \par\egroup 
  \egroup
  \setbox\TEMPBOX=\hbox to \TEMPDIMEN{%
    \hss
    \vbox{%
      \hsize=\wd\ItemBOX
      \ifvoid\tab@cap@box
      \else
        \noindent\unhbox\tab@cap@box
        \vskip 5.5pt plus 6pt%
      \fi
      \box\ItemBOX
      \iftable@txt
        \vskip 10pt%
        \eightpoint\nostb@ls{10pt}%
        \noindent\the\tab@txt@toks
        \global\table@txtfalse
      \fi
    }%
    \hss
  }%
  \ifsinglecol
    \if@here \midinsert\global\@herefalse\else \topinsert\fi
      \box\TEMPBOX
    \endinsert
  \else
    \global\setbox\ItemBOX=\box\TEMPBOX
    \JoinStack
    \Print{Processing source for table \the\ItemNUMBER}%
  \fi
}

\def\UnloadZoneA{%
\FirstZoneAtrue
 \Iteration=\z@
  \loop
   \ifnum\Iteration<\LengthOfStack
    \GetItemSTATUS{\Iteration}
    \ifnum\ItemSTATUS=\InZoneA
     \GetItemBOX{\Iteration}
     \ifFirstZoneA \vbox to \BodgeHeight{\vfil}%
     \FirstZoneAfalse\fi
     \unvbox\ItemBOX\ItemSep
     \LeaveStack{\Iteration}
     \else
     \advance\Iteration \@ne
   \fi
 \repeat
}

\def\UnloadZoneC{%
\Iteration=\z@
  \loop
   \ifnum\Iteration<\LengthOfStack
    \GetItemSTATUS{\Iteration}
    \ifnum\ItemSTATUS=\InZoneC
     \GetItemBOX{\Iteration}
     \ItemSep\unvbox\ItemBOX
     \LeaveStack{\Iteration}
     \else
     \advance\Iteration \@ne
   \fi
 \repeat
}


\def\ShowItem#1{
  {\GetItemAll{#1}
  \Print{\the#1:
  {TYPE=\ifnum\ItemTYPE=\Figure Figure\else Table\fi}
  {NUMBER=\the\ItemNUMBER}
  {SPAN=\ifnum\ItemSPAN=\Single Single\else Double\fi}
  {SIZE=\the\ItemSIZE}}}
}

\def\ShowStack{%
 \Print{}
 \Print{LengthOfStack = \the\LengthOfStack}
 \ifnum\LengthOfStack=\z@ \Print{Stack is empty}\fi
 \Iteration=\z@
 \loop
 \ifnum\Iteration<\LengthOfStack
  \ShowItem{\Iteration}
  \advance\Iteration \@ne
 \repeat
}

\def\B#1#2{%
\hbox{\vrule\kern-0.4pt\vbox to #2{%
\hrule width #1\vfill\hrule}\kern-0.4pt\vrule}
}


\newif\ifsinglecol   \singlecolfalse

\def\onecolumn{%
  \global\output={\singlecoloutput}%
  \global\hsize=\PageWidth
  \global\vsize=\PageHeight
  \global\ColumnWidth=\hsize
  \global\TextLeading=12pt
  \global\Leading=12
  \global\singlecoltrue
  \global\let\onecolumn=\relax
  \global\let\footnote=\sing@footnote
  \global\let\vfootnote=\sing@vfootnote
  \ninepoint 
  \message{(Single column)}%
}

\def\singlecoloutput{%
  \shipout\vbox{\PageHead\pagebody\PageFoot}%
  \advancepageno
  \ifplate@page
    \shipout\vbox{%
      \sp@pagetrue
      \def\sp@type{plate}%
      \global\plate@pagefalse
      \PageHead\vbox to \PageHeight{\unvbox\plt@box\vfil}\PageFoot%
    }%
    \message{[plate]}%
    \advancepageno
  \fi
  \ifnum\outputpenalty>-\@MM \else\dosupereject\fi%
}

\def\ItemSep{\vskip\ItemSepamount\relax}

\def\ItemSepbreak{\par\ifdim\lastskip<\ItemSepamount
  \removelastskip\penalty-200\ItemSep\fi%
}


\let\@@endinsert=\endinsert 

\def\endinsert{\egroup 
  \if@mid \dimen@\ht\z@ \advance\dimen@\dp\z@ \advance\dimen@12\p@
    \advance\dimen@\pagetotal \advance\dimen@-\pageshrink
    \ifdim\dimen@>\pagegoal\@midfalse\p@gefalse\fi\fi
  \if@mid \ItemSep\box\z@\ItemSepbreak
  \else\insert\topins{\penalty100 
    \splittopskip\z@skip
    \splitmaxdepth\maxdimen \floatingpenalty\z@
    \ifp@ge \dimen@\dp\z@
    \vbox to\vsize{\unvbox\z@\kern-\dimen@}
    \else \box\z@\nobreak\ItemSep\fi}\fi\endgroup%
}


\def\gobbleone#1{}
\def\gobbletwo#1#2{}
\let\footnote=\gobbletwo 
\let\vfootnote=\gobbleone

\def\sing@footnote#1{\let\@sf\empty 
  \ifhmode\edef\@sf{\spacefactor\the\spacefactor}\/\fi
  \hbox{$^{\hbox{\eightpoint #1}}$}\@sf\sing@vfootnote{#1}%
}

\def\sing@vfootnote#1{\insert\footins\bgroup\eightpoint\b@ls{9pt}%
  \interlinepenalty\interfootnotelinepenalty
  \splittopskip\ht\strutbox 
  \splitmaxdepth\dp\strutbox \floatingpenalty\@MM
  \leftskip\z@skip \rightskip\z@skip \spaceskip\z@skip \xspaceskip\z@skip
  \noindent $^{\scriptstyle\hbox{#1}}$\hskip 4pt%
    \footstrut\futurelet\next\fo@t%
}

\def\footnoterule{\kern-3\p@ \hrule height \z@ \kern 3\p@}

\skip\footins=19.5pt plus 12pt minus 1pt
\count\footins=1000
\dimen\footins=\maxdimen


\def\landscape{%
  \global\TEMPDIMEN=\PageWidth
  \global\PageWidth=\PageHeight
  \global\PageHeight=\TEMPDIMEN
  \global\let\landscape=\relax
  \onecolumn
  \message{(landscape)}%
  \raggedbottom
}


\output{%
  \ifLeftCOL
    \global\setbox\LeftBOX=\vbox to \ZoneBSize{\box255\unvbox\ZoneBBOX}%
    \global\LeftCOLfalse
    \MakeRightCol
  \else
    \setbox\RightBOX=\vbox to \ZoneBSize{\box255\unvbox\ZoneBBOX}%
    \setbox\MidBOX=\hbox{\box\LeftBOX\hskip\ColumnGap\box\RightBOX}%
    \setbox\PageBOX=\vbox to \PageHeight{%
      \UnloadZoneA\box\MidBOX\UnloadZoneC}%
    \shipout\vbox{\PageHead\box\PageBOX\PageFoot}%
    \advancepageno
    \ifplate@page
      \shipout\vbox{%
        \sp@pagetrue
        \def\sp@type{plate}%
        \global\plate@pagefalse
        \PageHead\vbox to \PageHeight{\unvbox\plt@box\vfil}\PageFoot%
      }%
      \message{[plate]}%
      \advancepageno
    \fi
    \global\LeftCOLtrue
    \CleanStack
    \MakePage
  \fi
}


\Warn{\start@mess}

\def\mnmacrosloaded{} 

\catcode `\@=12 


\fi
\catcode`\"=\active\let"=\"
\def\3{\ss }
%
\Autonumber  


\begintopmatter  

\title={Direct collisional simulation of 10 000 particles
   past core collapse }
  \author={R.~Spurzem$^1$ and S.~Aarseth$^2$}
\affiliation{$^1$ Inst. f"ur Astronomie und Astrophysik, Univ. Kiel,
    Olshausenstra\3e 40, 24098 Kiel, Germany}
\affiliation{$^2$ Institute of Astronomy, Madingley Rd., Cambridge
 CB3 0HA, U.K.}
\shortauthor{R. Spurzem and S. Aarseth}
\shorttitle{Direct $N=10\, 000$ simulation}
\acceptedline{Accepted, Received; in original form}
\abstract{A collisional $N$-body simulation using NBODY5 on a
single CRAY YMP processor is followed well into the post-collapse
regime. This is presently one of the largest particle
numbers of all such models
published, but some data for an even larger
$N$ produced by using special-purpose computers have
recently been presented.
In contrast to previous ensemble-averaged $N$-body
simulations the noise here is low enough to just compare this
one single run with the expectations from statistical models
based on the Fokker-Planck approximation. Agreement is as
good as could be expected for the case of the evolution
of the Lagrangian radii, radial and tangential
velocity dispersions and various core quantities.
We briefly discuss approximate models to understand number and
energy of escapers and the question of gravothermal
core oscillations; although the system exhibits
post-collapse oscillations they turn out to be
directly binary driven and we cannot prove
the existence of a gravothermal expansion
at this particle number. Finally in a detailed examination of the
wandering of the density centre we find in contrast to some previous
studies a clear long-time period of the order of about 14 half-mass
crossing times.
}
\keywords{stellar dynamics -- star clusters -- numerical methods
-- anisotropy}
\maketitle
\section{Introduction}
One of the grand challenges of theoretical astrophysics is to understand
the dynamics of globular star clusters. Apart from the intrinsic
interest of understanding the behaviour of large $N$-body systems, the
importance of the problem stems from its relation with current
observational research, and from the aim to study the
behaviour of even larger systems such as galactic nuclei. Without
understanding globular clusters by means of special tools for
investigating their evolution, the possibility of unravelling the
mysterious behaviour of galactic nuclei seems very remote. Unfortunately,
the direct simulation of such rich stellar systems as globular clusters
with star-by-star modelling is not yet possible. The gap between the
largest useful $N$-body models with $N$ being
of the order of $10^4$ particles
and the median globular
star cluster ($N \sim 5\times 10^5$) can only be bridged at present by
use of theory.  There are two main classes of theory: (i) Fokker-Planck
models (henceforth FP model),
which are based on the Boltzmann equation of the kinetic theory
of gases (Cohn, Hut \& Wise 1989, Murphy, Cohn \& Hut  1990),
and (ii) isotropic (Lynden-Bell \& Eggleton 1980,
Heggie 1984, Bettwieser \& Sugimoto 1984) and
anisotropic gaseous models (henceforth
AG models, Louis \& Spurzem 1991, Spurzem 1992, 1994),
which can be thought of as a set of
moment equations of the Fokker-Planck equation.

These simplified models are the only detailed models which are directly
applicable to large systems such as globular clusters. But their
simplicity stems from many approximations and assumptions which are
required in their formulation.  Examples are the assumption of
spherical symmetry, which
is inconsistent with the asymmetry of the galactic tidal field,
or statistical estimates of the
cross sections for the formation of close binaries by three-body
or dissipative (tidal) two-body encounters, and for their subsequent
gravitational interactions with field stars. Such processes play
a dominant role in reversing core collapse of globular clusters,
which otherwise would inevitably lead to a singular density profile
with infinite density at the centre (Lynden-Bell \& Wood 1968,
Bettwieser \& Sugimoto 1984,
Elson, Hut \& Inagaki  1987, Hut 1993).

Recent detailed comparisons
have established a good relationship between such simplified models
and direct $N$-body simulations, but for the even more idealized
case of a single mass system (identical individual mass in contrast
to real star clusters) and rather low particle numbers (Giersz \&
Spurzem 1994, henceforth GS, for $N=250$, $500$, $1000$ and $2000$; this paper
already contains some information on the
present $N=10000$ direct model, and also
Giersz \& Heggie 1994a, b, henceforth GHI, GHII). Only in one case has a
two-mass system been compared with direct $N$-body simulations
(Spurzem \& Takahashi 1995).
Unfortunately larger $N$
post-collapse star clusters cannnot be understood easily by
extrapolating the results for low $N$. Whereas two-body relaxation
scales with $N/\log \gamma N$ (with $\gamma \approx 0.11$), or in
other words, all $N$-body results {\sl before} core bounce
(the point where binary effects reverse core collapse) match
exactly if one scales the time with this factor
(GHI, GHII), the formation of three-body
and tidal two-body binaries introduces other timescales into
the problem, which do not scale with the same factor of $N$.

Therefore it remains very important to get
more data from direct $N$-body simulations with $N$ as high as allowed
by the present-day computational technology and software,
in order to strengthen confidence in the statistical models based
on the Fokker-Planck and other approximations. To follow for example
an $N=10000$ star cluster to core bounce
one needs to calculate some 900 initial
half-mass crossing times. To do this accurately a high-order integration
scheme with high intrinsic accuracy and no artificial softening of
the potential (as is very often used in so-called tree codes, Barnes
\& Hut 1986, for collisionless stellar dynamics, typically models
lasting some 10 or 20 initial crossing times) is necessary.

The required high accuracy integration scheme NBODY5 (Aarseth 1985)
includes a fourth order Adams-Bashforth-Moulton integrator
based on force interpolation polynomials at four past time
points. It includes as well a two- and more body regularization
technique (for two-body see Kustaanheimo \& Stiefel 1965)
to deal efficiently with close encounters and a two-level
timestep scheme (Ahmad-Cohen neighbour scheme, Ahmad \& Cohen 1973),
which updates the irregular force due to neighbour particles in
much shorter time intervals than the regular, less fluctuating
force exerted by the more distant particles.

NBODY5 has recently been improved to NBODY6, which uses a Hermite
interpolation for the force and its time derivative (based
on only two time points) and a hierarchical block time step scheme
(variation of time steps only allowed by a factor of 2 to keep
as many time steps commensurate as possible,
Makino \& Aarseth 1992). The latter is suitable for
implementation on parallel machines, and actually has been
implemented in a slightly advanced version NBODY6++ on parallel
computers (Spurzem 1995).

On the other hand special-purpose computers have been built to
calculate forces for gravitational $N$-body systems effectively
(Sugimoto et al. 1990). There are also so-called field programmable
arrays, hardware devices which can be programmed in a similar way
by hardware switches (Brown et al. 1992), which may be used for
direct $N$-body simulations in the near future (D. Merritt,
pers. communication). Such computers became increasingly available
in the past (for example the HARP-2 and HARP-3 computers
operational in Kiel and Cambridge, respectively) and
in autumn 1995 a Teraflop
machine consisting of many such HARP-boards has been presented
during the IAU symposium 174
in Tokyo (Hut \& Makino 1996),
which would be able to follow core collapse of a reasonably
realistic globular cluster. By using existing pieces of this
machine it already became possible to integrate systems larger than
$N=10^4$ past core bounce ($N=3.2\cdot 10^4$, Makino 1996)
and observe the onset of gravothermal oscillations.

In this paper we report on one of the largest published collisional
$N$-body simulations so far, which was performed during the last two
years on a single CPU of a CRAY YMP computer by using NBODY5.
Although it was not possible to detect gravothermal oscillations
without any doubt in that run, we demonstrate that $10^4$ is a
large enough particle number to exhibit a predictable evolution
of the system as a whole. Predictable means here, that there is
already reasonable agreement of this {\sl one} simulation with
the expectations from statistical models based on the Fokker-Planck
approximation. This increases confidence that it will be possible
to detect gravothermal oscillations in even larger simulations, which
are also predicted by the FP models. On the contrary,
for lower $N$, to check compatibility with theoretical models it is
necessary to average a sufficient number of statistically
independent $N$-body models (GHI, GHII, GS).
Single systems of $N=1000$, for example,
behave practically unpredictably because of the inherent chaotic
nature of the orbits in an encounter-dominated star cluster.
For large $N$, however, such chaotic behaviour of the individual
orbits is balanced by the decrease of the fluctuations, because the
relative energies of individual stellar encounters decrease compared to
the total energy of the system, as $N$ increases.

After some technical remarks for the model simulation (Sect. 2) we
report
in the following on the time evolution of some
relevant observables in the $N$-body simulation of an isolated, single
point mass, $N=10^4$ star cluster. These are mainly Lagrangian
radii, velocity dispersions and anisotropy, core parameters (Sect. 3.1), and
the results are compared with anisotropic gaseous models.
Escaper properties, the individual
evolution of hard binaries and their role for
the generation of post-collapse oscillations
are briefly examined (Sect. 3.2), and
the movement of the density centre is extracted from the data and
discussed in some detail (Sect. 3.3).

 \section{The simulation}

 A $10^4$-body realization (equal masses)
 of Plummer's model was integrated
 by using the standard NBODY5 scheme on one processor of a CRAY YMP
 machine for 2760 $N$-body time units. In the following we use standard
 $N$-body time units (Heggie \& Mathieu 1986), in which the
 total energy of Plummer's sphere is $E=-0.25$, $G=1$, and $M=1$
 ($G$, $M$, gravitational constant, total mass of the system,
 respectively). In
 such a system the individual body's mass is $m=1/N$
 for an equal mass cluster, where
 $N$ is the total particle number. The initial half-mass crossing
 time is $t_{\rm cr0} = 2\sqrt{2} \approx 2.828 $. Hence we
 have integrated the system over a little less than 1000 $t_{\rm cr0}$,
 which took about 1500 CPU hours on the
 CRAY, equivalent to approximately two months continuous running time.
 The accumulated error of the entire simulation in total energy
 was $\Delta E = 8.08 \cdot 10^{-3}$. The accuracy of the integration
 was checked by the criterion that the relative change in energy
 within a given time interval (0.5 $t_{\rm cr0}$) should
 be less than $10^{-5}$. Since NBODY5 contains the Ahmad-Cohen
 neighbour scheme each particle has two timesteps, a short one,
 after which the irregular force from the neighbours
 (typically less than $\sqrt{N}$ particles) is updated, and another regular
 timestep (which is usually about ten times larger than the
 irregular timestep), after which the
 total force exerted by all particles
 is updated, and changes in the neighbour
 list are maintained. Such a neighbour scheme saves a considerable
 amount of CPU time as compared to a scheme which at any step
 always calculates the full force, at least as one
 considers a non-parallel general purpose computer. Both the
 regular and irregular timestep are fully adaptive according
 to the indidvidual force variations on the particle
 (for details see Aarseth 1985). Our model performed in total
 approximately 13.5 billion irregular steps, 1 billion regular
 steps, and 500 million regularized steps for closely bound pairs,
 whose internal motion is integrated separately in regularized
 coordinates. This means on average each particle was moved
 1.35 million times (its regular force updated ten thousand times).
 A more quantitative information can be found on Figs. 1a, b, which
 show the CPU time used and the number of regular, irregular,
 and regularized (KS) steps per $N$-body time unit as a function
 of time (Fig. 1a) and the same data per time unit per particle
 (Fig. 1b). The fluctuations after $t=2300$ are due to the onset
 of the activity of extremely hard binaries in the core (core bounce
 is at $t=2380$, see below). These binaries suffer from strong
 superelastic scatterings with field stars in the core, so the
 timestep distribution of the field stars is affected itself.
 It is remarkable, however, that from the beginning until the
 core bounce, when there is a strong core-halo structure, the
 CPU time rates do not increase by more than half an order of
 magnitude, which is a proof that the two-level neighbour scheme of
 NBODY5 is indeed very well tailored to simulate centrally
 concentrated systems.
 \beginfigure{1}\nofloat
 \vskip 6cm
\includegraphics{fig1a.ps}
 \caption{{\bf Figure 1a.} CPU time in seconds, number of irregular,
 regular, and KS steps per $N$-body time unit, as a function of time.}
 \endfigure
 \beginfigure{2}\nofloat
 \vskip 6cm
\includegraphics{fig1b.ps}
 \caption{{\bf Figure 1b.} Same quantities as in Fig. 1a, but per
 $N$-body time unit per particle.}
 \endfigure

 \section{Results}
 \subsection{Spatial Evolution}

 In the following we compare our $N$-body results with those
 of a standard AG model (Louis \& Spurzem
 1991, Spurzem 1992, Spurzem 1994, Spurzem 1996),
 using for the latter those
 parameters which gave best agreement between all different
 kinds of star cluster models in previous
 work, namely $\lambda =0.4977$,
 $\lambda_{\rm A} = 0.1$ and $C_b = 90$. Here $\lambda$ is the
 numerical parameter of order unity scaling the heat flux,
 related to the standard $C$ constant in isotropic gaseous models
 (see e.g. Heggie \& Stephenson 1988) by
 $$\lambda = {27\over 10} \sqrt{\pi} C \ . $$
 Our $\lambda$ corresponds to the standard $C=0.104$ of Heggie
 \& Stephenson (1988), which gives best agreement between
 FP and AG models. $\lambda_{\rm A}$ is a
 numerical factor scaling the anisotropy decay time $T_{\rm A}$,
 defined such that the decay of
 anisotropy by two-body relaxation is
 $$(\delta A/\delta t)_{\rm rx} =
    - {A\over \lambda_{\rm A} T_{\rm A} }\ ,$$
 where $T_{\rm A}$ is the self-consistent
 anisotropy decay time for a
 Larson type anisotropic distribution function
 ($T_{\rm A}=10T/9$, with $T$ the standard two-body relaxation time).
 Our value of $\lambda_{\rm A}=0.1$ was fitted to averaged $N$-body
 simulations of $N\le 2000$.
 Finally $C_b$ is the numerical parameter scaling the energy generation
 by three-body binaries, whose standard value (Goodman 1987) is
 $C_b=90$, which we take here. Goodman \& Hut (1993)
 have recently argued that the formation rate of three-body
 binaries had been overestimated in previous work,
 so $C_b=75$ would ensue for their new results.
 But the still adopted value of $C_b=90$ is within the expected
 error range, for their derivations as well as the accuracy is
 concerned with which we can fix $C_b$ by comparison with the
 $N$-body simulation.
 The reader more interested in the
 details of how to define parameters
 like $\lambda $, $\lambda_{\rm A} $, and $C_b$ and how to select their
 values is referred to GS and the other cited papers above.
 \bigskip
 Fig. 2a compares the evolution of Lagrangian radii containing
 1-90 \% of the total mass between AG and our $N$-body model.
 Whereas the pre-collapse evolution agrees extremely
 well, and also the slope of the post-collapse expansion agrees
 fairly well, there is a difference in the collapse time itself,
 which is $t=2080$ for the gaseous model and $t=2380$ in the
 $N$-body system. So is there something wrong? One has to recall
 that the conductivity parameter $\lambda$ in the gaseous model
 was adjusted to match simultaneously the collapse time and slope of
 the Lagrangian radii evolution in pre-collapse for
 {\it averaged} $N$-body simulations ($N=250, 500, 1000$) and
 isotropic Fokker-Planck models (GS).
 In these averaged models there was always a considerable spread
 in collapse times of {\it individual} $N$-body models, as can
 be seen from Table 1. First we show in this table results obtained
 from various independent $N$-body calculations ($N\le 2000$), as
 e.g. the variance of Lagrangian radii for the initial model $\sigma_R$,
 minimum and maximum collapse time
 ($t_{\rm min}$ and $t_{\rm max}$) as measured in individual simulations,
 the average collapse time $t_{\rm cc}$ for all simulations of a given particle
 number, the variance of the core collapse time $\sigma_{\rm cc}$ and
 its relation to the collapse time itself. For more details about
 the averaging process and the individual simulations see GHI and GHII,
 from whose models these data were taken.
 For all quantities except $t_{\rm min}$ and $t_{\rm max}$ there is
 an expected scaling with $N$ denoted in Table 1; we found it in
 general agreement with the measured values, although there is some
 tendency for mismatch with such scaling for low $N$ ($N\le 500$). In
 particular the scaling law for the collapse time stems from the
 assumption that it is two-body relaxation (by taking a Coulomb
 logarithm factor of $\gamma = 0.11$, see GHI, GHII)
 which dominates the evolution to core
 collapse.

 We have used these scaling laws to extrapolate
 quantities for $N=10000$ and $N=20000$, which cannot be measured yet,
 because there are not enough individual $N$-body simulations (values
 for $N=1000$ were used to extrapolate for $N=10000$, and those of
 $N=2000$ to extrapolate to $N=20000$). We find for 10000 particles
 an extrapolated average collapse time of $t_{\rm cc} = 2215$. Note that
 even this value is not very accurate, because the {\sl measured}
 average collapse times do not exactly obey the assumed relation.
 But if we take this value for the time being as a fiducial value
 it follows, that our individual $N$-body simulation has a collapse
 time 165 time units (1.14 $\sigma$) {\sl larger} than the average, and
 gaseous and Fokker-Planck models have a collapse time 135 time units
 (0.93 $\sigma $) {\sl smaller} than the expected average collapse time.

 We stress these numbers here in detail, because recently it has
 been claimed
 that the collapse time in anisotropic systems could be somewhat
 larger as in the isotropic
 case, as suggested by new 2D Fokker-Planck models of Takahashi
 (1995, 1996), and previously by higher order moment models of Louis (1990).
 However, at this present particle number of $N=10000$ it seems very
 difficult to distinguish such a variation from the intrinsic
 uncertainties in determining the core collapse time. There are
 several factors which influence measured collapse times, as e.g.
 in $N$-body models the number of independent simulations, and in
 gaseous and Fokker-Planck models variation of quantities like $t_{b0}$ and
 $C_b$ (starting time and strength of binary energy generation, see
 GS for their definition, and their Figs. 4 and 5 for their influence
 on collapse times) and the factor $\gamma$ used in the Coulomb logarithm
 could always cause a few per cent change in the collapse time. For example,
 $\gamma =0.11$ has been adopted as the best value from GHI and GHII, who
 considered only $N\le 2000$; a smaller $\gamma$ could always cause
 a larger collapse time. Hence we conclude here that from the present
 data there is some suspicion that the collapse time in $N$-body
 simulations might be longer than expected from anisotropic gaseous
 and isotropic Fokker-Planck models, however, we
 consider the difference of just
 one $\sigma$ as not yet very convincing. Note that the free parameter
 in the conductivity of the {\sl anisotropic} gaseous model was scaled
 to the same collapse time as the {\sl isotropic} Fokker-Planck model (GS).
 Thus if one came to the conclusion that anisotropic 2D Fokker-Planck
 models and naturally anisotropic $N$-body systems have a slightly
 longer collapse time, this just would mean that the anisotropic
 gaseous model had to be rescaled in its conductivity parameter
 $\lambda$, now with respect to the 2D Fokker-Planck model.
 To judge about these questions has to be postponed to the future,
 when $N$-body simulations of large particle numbers ($N\ge 10000$)
 become available in a larger number.

\begintable{1}
\caption{{\bf Table 1.} Radial variance of initial models $\sigma_R$,
minimum and maximum core collapse times $t_{\rm min}$ and $t_{\rm max}$,
average core collapse time $t_{\rm cc}$, variance
of core collapse time $\sigma_{\rm cc}$, and
the ratio $\sigma_{\rm cc}/t_{\rm cc}$ as a function of particle
number $N$. Values with a *-symbol are estimated from the scaling
laws indicated in the last lines. All data for $N\le 2000$ kindly
supplied by M. Giersz (priv. comm.)}
\halign{
 \hfil  # & \hfil # & \hfil # &\hfil # &\hfil # &\hfil # &\hfil # \cr
     N    &  $\sigma_R$ & $t_{\rm min}$ & $t_{\rm max}$ & $t_{\rm cc}$ &
     $\sigma_{\rm cc}$   & $\sigma_{\rm cc}/t_{\rm cc}$ \cr
 250      & 2.54E-2 &  23     & 144    & 80     & 25     &  0.3125\cr
 500      & 2.02E-2 &  75     & 232    & 156    & 34     &  0.2179\cr
 1000     & 1.62E-2 & 255     &  450   & 330    & 46     &  0.1394\cr
 2000     & 1.17E-2 & 505     &  740   & 626    & 59     &  0.0942\cr
 10000    & 5.12E-3*&  ?      &  ?     & 2215*  & 145*   &  0.0655\cr
 20000    & 3.70E-3*&  ?      &  ?     & 4387*  & 187*   &  0.0426\cr
\noalign{\smallskip}
Scaling   &${1\over\sqrt{N}}$ &     &   &${N\over\log(\gamma N)}$ &
                                       $\sqrt{N}$ &
     ${\log(\gamma N)\over\sqrt{N}}$\cr}
\endtable

In Fig. 2a one can see at a time much later than
core bounce ($t\approx 4000$) that the
gaseous model becomes unstable to gravothermal oscillations at
very late times (compare for gravothermal oscillations Bettwieser
\& Sugimoto 1984, Goodman 1987, Breeden, Cohn \& Hut 1994).
Since the instability at $N=10 000$ is not very strong
(particle number not much above the critical number 7000 for
gravothermal oscillations in the equal mass system) the
oscillations start so late, at a time, which our $N$-body
simulation did not yet reach. It exhibits, however, some
kind of oscillations directly after core bounce. But, as will
become clear during the discussion of the binary effects,
we cannot establish firmly that these oscillations are
gravothermal, in contrast to just directly driven by
binary energy. Gravothermal oscillations (or expansion)
should continue even after binary energy production has ceased.

\beginfigure{3}\nofloat
\vskip 6cm
\includegraphics{fig2a.ps}
\caption{{\bf Figure 2a.} Lagrangian radii containing the indicated
fraction of total mass as a function of time for the $10^4$-body
simulation (fluctuating curves) and the standard anisotropic
gaseous model (smooth curves).}
\endfigure
\beginfigure{4}\nofloat
\vskip 6cm
\includegraphics{fig2b.ps}
\caption{{\bf Figure 2b.} Lagrangian radii as in Fig. 2a, but
here our $10^4$-body simulation (time rescaled according to
main text) compared to an averaged $N=1000$ simulation of
GHI and GHII.}
\endfigure

To relate our single $N=10 000$-body model with the
lower $N$ cases of GHI and GHII we have scaled its time
coordinate by $C(N_0)/C(N)$, where
$C(x):=x/\log(\gamma x)$ ($\gamma = 0.11$) for $N=10^4$ and
$N_0=1000$, and plotted the data for the Lagrangian radii together
with those of an averaged $N=1000$ body calculation, kindly made
available by M. Giersz, in Fig. 2b. Such scaling would occur
if standard two-body relaxation is the dominant force of evolution,
which in fact can be clearly seen in the plot.
The separation
of the curves for different $N$ at core bounce is due to
the three-body activity which sets in at different times
and scales with a different power of $N$.

\beginfigure{5}\nofloat
\vskip 6cm
\includegraphics{fig3.ps}
\caption{{\bf Figure 3.} Lagrangian radii as in Fig. 2a, but now
an enlarged time segmented near core bounce. Gaseous model rescaled
according to text in order to compare the post-collapse slopes
of both models. The times of prominent hard binary events are
marked, as there are binary escapers (dots) and
strog hardening events of binaries by three-body encounters
(stars).}
\endfigure

In order to compare the post-collapse evolution with
the gaseous model it is better to scale the times of the gaseous
model to the exact collapse time ($t=2380$) of the $N$-body simulation.
The result on an enlarged timescale can be seen on Fig. 3. To get
Fig. 3 we have multiplied the time and all radii with a free factor,
so as to match the collapse time and the vertical position of the
gaseous model and $N$-body curves.
Consistently with the previous
results of GHI, GHII, and GS we find that core bounce in the
$N$-body model is deeper in the inner shells and that the intermediate
zones of the gaseous model expand faster, whereas the outermost shells
lag behind as compared to the $N$-body model. We think this is due
to a non-local energy transport by high-velocity
stars on radial orbits, which are reaction products
from superelastic binary encounters in the core, as elaborated also
in the previously cited papers.
In the gaseous model energy can
only be transported locally by heat conduction with the
temperature gradient; thus the energy created by binaries, which
is of the correct magnitude (see below) causes more expansion
in the adjacent, intermediate shells relative to the $N$-body model.
The importance of high-velocity stars carrying energy from superelastic
encounters quickly out of the core was first realized by
Aarseth (1974) and further elaborated on by Makino
\& Sugimoto (1987), who denoted them as suprathermal particles.

\beginfigure{6}\nofloat
\vskip 6cm
\includegraphics{fig4a.ps}
\caption{{\bf Figure 4a.} Radial 1D velocity dispersion averaged
between the indicated Lagrangian radii as
a function of time. Comparison between
$N$-body (fluctuating curves) and gaseous model (smooth curves).}
\endfigure
\beginfigure{7}\nofloat
\vskip 6cm
\includegraphics{fig4b.ps}
\caption{{\bf Figure 4b.} As Fig. 4a, but for an enlarged time
section near core bounce.}
\endfigure
\beginfigure{8}\nofloat
\vskip 6cm
\includegraphics{fig4c.ps}
\caption{{\bf Figure 4c.} As Fig. 4a, but for the tangential 1D
velocity dispersion.}
\endfigure
\beginfigure{9}\nofloat
\vskip 6cm
\includegraphics{fig4d.ps}
\caption{{\bf Figure 4d.} As Fig. 4b, but for the tangential 1D
velocity dispersion.}
\endfigure
\beginfigure{10}\nofloat
\vskip 6cm
\includegraphics{fig4e.ps}
\caption{{\bf Figure 4e.} As Fig. 4a, but for the anisotropy
defined as $A=2-2\sigma_t^2/\sigma_r^2$.}
\endfigure

Figs. 4a-d show the radial and tangential velocity dispersions,
each for the entire evolutionary time (4a, 4c) and
for an enlarged time segment around core bounce (4b, 4d). In most
cases the general agreement between the gaseous model and
our single $N$-body
simulation is striking. However, there are some features
near and after core bounce worth
discussing in more detail. First,
the effect that intermediate shells (e.g. 10-20 \% of total mass)
expand a little faster than the $N$-body system, as was
detected for the Lagrangian radii, corresponds to a smaller
``temperature'' (velocity dispersion) there.
On the other hand, further outside, where the $N$-body
system expands more rapidly, it is somewhat cooler than the AG model.
In general one could say that the large scale Figs. 4a and 4c support
our claim of the generally fair agreement between $N$-body and
AG models, but Figs. 4b and 4d presenting a
zoom in on core bounce and post-collapse give a clearer picture of
what are the remaining differences. How about the anisotropy
$A=2-2\sigma_t^2/\sigma_r^2$ itself? In principle it is redundant, all
information is already contained in the previous plots. However,
the analysis of $A$ enhances the remaining differences between
the models, so we will discuss it here as well with the help of
Fig. 4e. For the innermost
Lagrangian radii (say up to 30 \%) it is not useful to discuss $A$,
because it is fluctuating strongly; comparisons with the
AG models have to rely on
the velocity dispersions as above. Note, however,
the good agreement between both models
for the anisotropy
averaged between 30 and 40 \% of total mass (lowermost curves
of Fig. 4e), especially
in the reduction of anisotropy after core bounce. Surprisingly,
there is not
such agreement for the larger radii. For 40-50 \%
the decrease of $A$ in post-collapse is only visible in the AG model,
the anisotropy in the $N$-body model increases further. More
outside, the disagreement between both models starts
even earlier. This is consistent with the idea of suprathermal
particles, moving on radially elongated orbits towards the halo,
not being accounted for in the gaseous model. As for the outermost
shells of the system one has also to take into account a difference
in the boundary conditions (the $N$-body model is open and may
expand freely into space, the AG model is confined by an
adiabatic wall at 100 \% of total mass, so the 100 \% Lagrangian
radius cannot expand, which it does in the $N$-body model).

Figs. 5a-c depict the time development of
the central potential, Cohn's $x$ value ($x=\Phi_c/3 \sigma_c^2$, where
$\sigma_c$ is the core velocity dispersion), and the number of
core particles.
All of the data in Figs. 5
are scaled in time as in Fig. 2b,
in order to compare with the data of GHI and
GHII. In some cases the figure caption notes that the data
were smoothed, which means the high frequency fluctuations
have been cut out by the ``Numerical Recipes'' routine SMOOFT, which
uses Fourier transforms to do that. The window for smoothing
was about 30 data points, which are 45 time units. In Figs. 5a and c
once more a very good agreement of a {\sl single} $N=10^4$ model
with the {\sl average} $N=1000$ data is visible for the pre-collapse
phase. During core bounce the single $N=10^4$ simulation exhibits
strong fluctuations (very deep central potential, correlated with an
extremely small core particle number), which presumably would be
smoothed out if we would be able to average the results for different
independent models, as in the $N=1000$ case.

\beginfigure{11}\nofloat
\vskip 6cm
\includegraphics{fig5a.ps}
\caption{{\bf Figure 5a.} Central Potential as a function of scaled
$N$-body time, see text. Compared with a smoothed curve obtained
from average $N=1000$ data.}
\endfigure
\beginfigure{12}\nofloat
\vskip 6cm
\includegraphics{fig5b.ps}
\caption{{\bf Figure 5b.} Cohn's $x$ as a function of scaled
$N$-body time. Smoothed in time, compared with smoothed data
from average $N=1000$ simulation.}
\endfigure
\beginfigure{13}\nofloat
\vskip 6cm
\includegraphics{fig5c.ps}
\caption{{\bf Figure 5c.} Core particle number as a function of scaled
$N$-body time. Compared with data of average lower $N$ simulations.}
\endfigure

\beginfigure{14}\nofloat
\vskip 6cm
\includegraphics{fig6.ps}
\caption{{\bf Figure 6.} Individual particle's potentials for different
times in pre- and post-collapse as indicated.}
\endfigure

It is very instructive to look at a graph just showing as tiny dots
values of the potential at the individual particle locations
as a function of radius for different times in pre- and post-collapse
(Fig. 6). This gives a good feeling how smooth the particle
distribution already is for a total particle number of $10 000$,
and moreover, it unambigously clarifys that core bounce, the
time with the deepest central potential, occurred at $t=2380$
(labelled 886 in the plot),
which can also be seen in Fig. 5a (but note the rescaled time).

\subsection{Escapers and binaries}

At the present endpoint of our simulation a total of 360
particles had escaped (they were removed from
the calculation when speeding with more than the escape velocity
from the entire cluster and being at a radius larger than
ten times the initial scaling radius
of Plummer's model). Fig. 7a,b show the cumulative
particle and energy loss. Similarly to GHI we tried to
approximate the $N$-body results by rescaling H\'enon's (1965)
escape rate (isotropic Plummer sphere)
multiplicatively for the effects of
evolution and anisotropy (see Figs. 9 and 11 of GHI). The effect
of evolution was determined by directly calculating the escape rate of
an isotropic Fokker-Planck model with $N=10000$ particles and
standard parameters ($\gamma = 0.11$ for the Coulomb logarithm,
$C_b=90$ for the energy generation), see for reference Eq. 2 of GHI.
As a model for the effect of anisotropy Dejonghe's (1987) anisotropic
Plummer models were taken, which have a dimensionless parameter $q$.
$q$ can be determined by taking the anisotropy $A$ at a certain
radius $r$ of the $N$-body model by $q=A/(1+1/r^2)$. GHI took the anisotropy
at the Lagrangian radius containing 75 \% of total mass, here we
try to bracket the $N$-body results by computing $q$ and subsequently
scaled escape rates for both the 75 \% and 90 \% Lagrangian radius
(see plots). Since we were able to distinguish every single escaper
in our $N$-body calculation, it was possible to remove those escapers
by number and energy, which are due to three-body
encounters (scatterings with hard binaries, in the
following denoted as ``hard escapers''), defined by the condition
that their energy of escape is larger than 3 kT. Such reduced
particle and energy loss is labelled by N-BIN in the plots.

{}From the figures we conclude the following: the effect of hard
escapers in the $N$-body rates of escape is small for the
particle number and in pre-collapse, but dominant for the energy
of escapers in core bounce and beyond. A fair approximation of the
particle and energy loss in the $N$-body system is reached by
rescaling H\'enon's escape rate for the effect of evolution and
anisotropy, provided the energy of the hard escapers is subtracted first from
the $N$-body data. The effect of evolution visible in the curves
labelled FP-I is much less significant than that of the anisotropy.
However, since Dejonghe's models are only a very approximate
model it is not a priori clear, how to fit the best $q$ parameter
to an individual $N$-body system. To illustrate this uncertainty
we have used two different Lagrangian radii (containing 75 \% and
90 \% of total mass, with correspondingly different anisotropies
and thus varying $q$ parameter for the same time). The escape rate
of particles seems to be best approximated by using a $q$-value derived
from the 90 \% Lagrangian radius (Fig. 7a), whereas the energy loss
rate is bracketed by the two choices (Fig. 7b). It follows that it is not
possible to find a simultaneous best fit to both data; we think
that this is due to the fact that Dejonghe's models are not an accurate
representation of the real $N$-body distribution. Note that
we had to rescale here once more the $N$-body time
to the same collapse time as in the Fokker-Planck model
($t=2080$, see above Fig. 2a and its discussion) used
for comparison, in order to get a useful comparison of both data.

\beginfigure{15}\nofloat
\vskip 6cm
\includegraphics{fig7a.ps}
\caption{{\bf Figure 7a.} Cumulative particle loss due to escapers
as a function of scaled $N$-body time (scaled to the collapse time
of an isotropic Fokker-Planck model). Curves labelled N and N-BIN denote
the $N$-body data with and without hard escapers (see main text); the
data labelled FP-I show the increase of H\'enon's (1975)
standard escape rate (at $t=0$) due to the
evolutionary effect; the curves
labelled A-90 and A-75 show the multiplicative combination
of the evolutionary effect and another correction for
anisotropy evaluated by using Dejonghe's (1987) anisotropic Plummer model
with a $q$-value derived from the anisotropy at the 75 \% or 90 \%
Lagrangian radius, respectively.}
\endfigure
\beginfigure{16}\nofloat
\vskip 6cm
\includegraphics{fig7b.ps}
\caption{{\bf Figure 7b.} As Fig. 7a, but for the
energy of the escapers.}
\endfigure

We started initially without any binaries. At $t=1200$ a
transient mildly hard binary formed, whose effect can be seen
in a slight dip of the $N$-body Lagrangian radii against
the gaseous model data in Fig. 2a. Such very early transient
binary activity may be a reason, why our model system has a
collapse time, which is at more than one $\sigma$ above the
expected average value. The really hard binaries then formed
shortly before core bounce.
This can be seen on Figs. 8a, b, where the high energy ($E\gg 1 kT$)
escapers occur just at core bounce, which is the time of
formation of hard binaries, as is shown in Fig. 8c. Fig. 8b
is an enlargement from the data of Fig. 8a; we see
that the binary activity reduces to a significantly lower
level again at $t=2600$, since we interpret any escaper
with more than $1kT$ as due to an interaction with a hard
binary. It is interesting to note, that near the times $t=2400$, $2500$,
and $2600$ there a crowding of high energy escapers in
Fig. 8a. We relate this to the spikes in the velocity dispersions
of the outer Lagrangian shells occurring in Figs. 4 just near
the same time, only for the radial
velocity dispersion, not for the
tangential one. This supports the idea, that the high-energy reaction
products of the strong three-body encounters move radially outwards
and generate the spikes in $\sigma_r$. Moreover, these three times
can be identified as periods of strong binary activity in Fig. 8c
(binaries exceeding a binding energy of 200 in $N$-body units). If one
then looks at Fig. 3 these times approximately correspond to expansion
phases of the Lagrangian radii; after the binary activity has ceased,
the expansion does not continue, on the contrary, the system starts
to contract again. Note, that after $t=2600$ there is also no
strong binary activity, because the hardest binary, although it is
still contained in the system (not yet escaped), moves on an elongated
orbit after a recoil. Hence it has a small probability for three-body
encounters. From all this evidence we conclude that the post-collapse
oscillations of the $N$-body system are binary driven and not
gravothermal. If they were gravothermal, expansion should continue
after the binary energy generation has ceased. Moreover, as
already Bettwieser \& Sugimoto (1984) stated there should be
a radial temperature (i.e. velocity dispersion) inversion in
the case of a gravothermal expansion, which
we could not observe with a sufficient statistical significance in
our $N$-body data. But on the other hand, there is a steady
expansion consistent with the gaseous model in post-collapse
superposed on the binary driven oscillations. So we would not exclude
that on a secular timescale our $N$-body system will undergo a
gravothermal expansion, presumably with a temperature inversion too
small to be extracted from the noisy data.

\beginfigure{17}\nofloat
\vskip 6cm
\includegraphics{fig8a.ps}
\caption{{\bf Figure 8a.} Time at which stars escape as a
function of their energy in $kT$, where $1kT = \sigma_c^2$,
with the 1D core velocity dispersion $\sigma_c$.}
\endfigure
\beginfigure{18}\nofloat
\vskip 6cm
\includegraphics{fig8b.ps}
\caption{{\bf Figure 8b.} As Fig. 8a, but with an enlarged
time axis to see the decrease in binary activity after core
bounce for $t> 2600$.}
\endfigure
\beginfigure{19}\nofloat
\vskip 6cm
\includegraphics{fig8c.ps}
\caption{{\bf Figure 8c.} Individual fate of binaries close
to core bounce; binding energy as a function of time. At some
times labels are plotted, which are the names of the stars
in the binary and can be used to identify them in the following
plot. Three thick vertical line segments show the magnitude of
an increase in binding by energy by 40\% from energy 20, 100, and
300, respectively, to be compared with the actual energy changes.
Binary escapers and exchange reactions of the binaries are indicated.}
\endfigure
\beginfigure{20}\nofloat
\vskip 6cm
\includegraphics{fig8d.ps}
\caption{{\bf Figure 8d.} Radial position of the hard binaries
of the previous figure over the same time span. Note that the core
radius is much smaller than one during this entire period. In some
cases by comparison with Fig. 8c the event can be identified,
which kicked a binary out of the core or even out of the cluster
(lines escaping to the top belonging to escaping binaries).}
\endfigure
\beginfigure{21}\nofloat
\vskip 6cm
\includegraphics{fig8e.ps}
\caption{{\bf Figure 8e.} Energy balance as a function of time
near core bounce. Labelled are the curves for the binding energy
of the binaries still inside the cluster ($E_{\rm bin}$),
the cumulative translational energy of single escapers ($E_{\rm esc}$),
the cumulative binding energy carried away by escaping binaries
($E_{\rm escbin}$), and the cumulative energy due to three-body
binaries produced in a standard AG model
($E_{\rm bingas}$). The core collapse time of the gaseous model
was scaled to the actual collapse time of the $N$-body model, see
text.}
\endfigure

In Fig. 8c
the fate of the hard binaries in the system can be followed
individually, e.g. one can estimate by eye that the rate
of change of the binding energy of a hard binary is related
to its binding energy $x$; in an encounter binaries with large $x$
suffer a change in binding energy of the same order
as $x$. This is consistent with the early experimental (Hills 1975)
and theoretical (Heggie 1975) predictions, that on average
the energy change $\Delta x $ during a superelastic
scattering should be $ \Delta x = \eta x$ ($\eta = 0.4$). GHII argue that
an experimental value of $\eta$, found by their statistics of a large
number of $N$-body simulations for $N\le 2000$, is about 0.2,
in better agreement with more recent predictions by Spitzer (1987) and
scattering experiments by Heggie \& Hut (1993). Since,
however, in contrast to the averaged $N$-body simulations of
GHI and GHII, we only have a rather small number of strong
encounters, i.e. a poor statistics, we will not elaborate further
on this point here. We think our results are at least consistent
with a value of $\eta$ between 0.2 and 0.4;
a change with
$\eta = 0.4$ is
indicated in
the figure as a thick vertical bar
for three energies $x=20, 100, 300$.
One may wonder
why in one case at about $t=2570$ the binding energy
of a binary decreases so dramatically. This is due to
the formation of a hierarchical triple, which was unfortunately
not included in our binary list. Later, its binding
energy occurs again in the most strongly bound binary,
after dissolution of the triple. Fig. 8d
gives an optical impression of the orbital motion of the
centres of masses of hard binaries,
their recoils by superelastic scatterings to very large
radii, and subsequent shrinking again by dynamical friction to
the core. Especially the motion of the hardest binary consisting of
the particles \# 4805 and 5745 (see Fig. 8c, its
binding energy amounts to several hundred $kT$) reaches far into the
halo after an energectic recoil. Thus one
can understand that exotic products of binary evolution
(like soft X-ray binaries or blue stragglers, see for the
first e.g. Rappaport et al. 1994) may be
found in the outskirts of a cluster.

As a final remark we want to demonstrate with Fig. 8e how
the cumulative binary energy generation in the $N$-body system
compares to the steady energy input modelling the binary
heating in the gaseous model. One can see that the energy
liberated in the gaseous model is of the same order as the
binding energy of the hard binaries still in the $N$-body system,
and also of the same order as the binding energy carried
away by escaping binaries.

\subsection{Movement of the density centre}

A very interesting and not yet completely understood feature
of a real $N$-body system is the oscillations of its density
centre and also other quantities, like the Lagrangian radii.
To define the density centre we first compute for each
particle $i$ a neighbour density $\rho_i$,
according to Casertano \& Hut (1985)
by the mass inside $r_6$ (excluding the
test particle itself), where $r_6$ is the distance of the
sixth nearest neighbour. Then by
$${\bf r}_d = \sum_{(i)} \rho_i {\bf r}_i \bigr/
	      \sum_{(i)} \rho_i  $$
we define a density centre for a special subset of particles
(${\bf r}_d$, ${\bf r}_i$ are the position vectors of the density
centre and the particle $i$, resp.). For the summation only the
inner parts of the system should be included, in order to exclude
strong biases by asphericities in the outermost parts of
the system (large weight due to large $\vert{\bf r}_i\vert$). On the
other hand selecting the core particles or some fraction of them
(based on the previous core radius, which was also determined in
a similar way as the density centre) proved to produce extreme
fluctuations of the density centre in the late phases of core collapse.
So we chose a procedure suggested by McMillan, Hut \& Makino (1990), to
take all particles whose $\rho_i$ is some fraction (actually $1/20$)
of the maximum $\rho_i$. This yielded the data for the density
centre of Figs. 9a-c. In recent years there were detailed
examinations of core wandering and oscillations by Heggie, Inagaki
\& McMillan (1994) and Sweatman (1993). Our data are different from
their data in several respects. Firstly, we cover a much larger
time, however, do not resolve the small timescales as they did.
We cover approximately 1200 N-body time units with a constant time
interval of roughly half an initial mass crossing time ($\Delta t = 1.4$).
In principle our dataset is even larger, but unfortunately we
did not always keep a constant time interval for the data collection,
which turns out to be essential for the subsequent analysis.
Secondly, we are not able to resolve the noise at high frequencies,
as was the aim of Heggie, Inagaki \& McMillan (1994). In fact, our
data become significant just at frequencies smaller than 0.1
(see their Fig. 7), where their statistics
becomes poor. Also Sweatman (1993), who did a very thorough
examination, which we closely follow here in our data processing,
was limited to periods smaller than 10 (frequencies larger than 0.1),
due to the duration of his simulation.

We first present the original data of the density centre in
three different scales in Figs. 9a-c, together with the
centre of mass data for the entire system. As noted already
by the cited authors and also by Makino \& Sugimoto (1987)
there are variations on different timescales, beginning from
very abrupt changes on times comparable to the dynamic time
up to larger changes on timescales of the order of several
crossing times. We will now inspect the time variability
of the data in more detail.

In Figs. 9b, c a third
(thick) curve is plotted, which is the density centre data
${\bf r}_{sd}$ smoothed
by the SMOOFT procedure of ``Numerical Recipes'' with a window of
50 data points (we varied the window to 30 and 100
data points without any
significant change in the following results). The reason is that
for a proper autocorrelation and frequency analysis we need data
whose average is zero. Sweatman (1993) took the centre of mass
as a reference point, but as one can see in Fig. 9a for a long
time simulation, where many, partly high energy, escapers are removed
and the outermost parts of the system are not nicely symmetric anymore,
there is a systematic shift between the centre of mass and the
density centre. Therefore we need the smoothed data as reference,
such that we can compute a quantity
${\bf \delta R} = {\bf r}_{sd} - {\bf r}_d$, whose time average is zero.
Therefore we cannot detect any periods in the data which are
{\sl larger} than the smoothing window, so we checked
that our results are independent of the size of it. With
$\delta R = \vert{\bf \delta R}\vert$ we performed an autocorrelation
analysis identical to that described by Sweatman (1993) in his
chapter 3.3. The striking result is in Fig. 9d, which exhibits
a dominant period of the order of 40 time units,
which is about 14 half-mass crossing times. To check this result
independently we took the data for $\delta R$ and performed
numerically a discrete Fourier transform (by use of the
Four-function of Mathematica). The resulting power spectrum as
a function of period is displayed in Fig. 9e - again there is
a clear maximum near periods of the order of 50, one could distinguish
two subpeaks at about 40 and 60.

\beginfigure{22}\nofloat
\vskip 6cm
\includegraphics{fig9a.ps}
\caption{{\bf Figure 9a.} Modulus of the vector of the density
centre and the centre of mass of the entire cluster as a function
of time.}
\endfigure
\beginfigure{23}\nofloat
\vskip 6cm
\includegraphics{fig9b.ps}
\caption{{\bf Figure 9b.} As Figure 9a, with enlarged time axis,
and added the thick curve for the smoothed density centre variations,
which serves as a reference for the autocorrelation analysis,
see text.}
\endfigure
\beginfigure{24}\nofloat
\vskip 6cm
\includegraphics{fig9c.ps}
\caption{{\bf Figure 9c.} As Figure 9a, but even more zoomed in
on the time axis, to see the limiting resolution of our data at
small time scales.}
\endfigure
\beginfigure{25}\nofloat
\vskip 6cm
\includegraphics{fig9d.ps}
\caption{{\bf Figure 9d.} Autocorrelation Function explained in the
text as a function of displacement (in time) $\lambda $ (in $N$-body
time units).}
\endfigure
\beginfigure{26}\nofloat
\vskip 6cm
\includegraphics{fig9e.ps}
\caption{{\bf Figure 9e.} Power spectrum of the differences between
the density centre data in Figs. 9a-c and their smoothed counterpart
(see Figs. 9bc), as a function of period in $N$-body time units.}
\endfigure

 \section{Conclusions and discussion}

 We have performed a very long (in terms of computer time as
 well as of physical time) simulation of a collisional $N=10^4$
 particle system well into the post core collapse evolutionary
 phase by using the standard NBODY5 integrator (Aarseth 1985).
 As initial model a Plummer sphere with no binaries and equal
 masses was selected.
 In contrast to previous averaged $N$-body simulations
 (Giersz \& Heggie 1994a, b, GHI, GHII; Giersz \& Spurzem 1994, GS)
 this single model already compares well with expectations of
 gas dynamical models based on the Fokker-Planck approximation.
 In particular the evolution until core collapse is
 entirely dominated by standard two-body relaxation,
 in particular Lagrangian radii, velocity dispersions and
 anisotropy compare fairly well with the other, statistical models,
 although there are some disagreements, especially in post-collapse
 and for the outer halo regions, which we think are due to the poor
 representation of suprathermal particles, originating
 from superelastic scatterings with hard binaries, in the gaseous model.
 Anisotropy is also the dominant factor determining the time
 evolution of the escape rate, which in deep core collapse is
 much larger than H\'enon's (1965) standard estimate. Within
 some uncertainty the effect of the anisotropy on the escape rate
 can be modelled by using one of Dejonghe's (1987) anisotropic
 Plummer models, provided the energy of escapers from hard three
 body encounters is not taken into account.

 Thereafter the core bounces due to the energy generation
 by superelastic scatterings between single stars and binaries.
 The energy generation is in fair accord with statistical predictions
 by Heggie (1975) and Hut (1993). From our simulation we
 see that individual hard binaries are sometimes ejected by
 recoil effects very far from the core after encounters.
 Since in our simulation we are able to follow the individual
 fate of each binary, we can correlate strong binary activity
 with expansion phases of the whole system. Since such expansion
 does not continue after the binary activity ceases, the post-collapse
 oscillations present in our model are {\sl not} gravothermal in
 nature. Although theoretically 10000 particles should be enough
 to observe gravothermal oscillations (but they set in long
 time after core bounce in the gaseous model), the fluctuations
 induced by individual strong three-body encounters suppress them
 to a non-observable level. Nevertheless the $N$-body expands on
 a secular timescale as prescribed by the gaseous model, so it
 cannot be excluded that there is a gravothermal expansion with
 superimposed binary-driven oscillations. In recent simulations
 (Makino 1996) with even
 larger particle number (32k particles) the onset of gravothermal
 oscillations including a temperature inversion has been observed,
 which would mean that for increasing $N$ the gravothermal
 evolution as predicted by gaseous and Fokker-Planck models will
 dominate the evolution as compared to the stochastic variations
 directly induced by binary activity. For a particular case,
 namely the fluctuations of core collapse time in an
 individual $N$-body simulation, we could show that it will
 decrease relative to the collapse time itself with larger $N$.

 Most features of our single $N=10^4$ model
 are consistent with expectations from statistical
 (gaseous or Fokker-Planck models), although our collapse time is
 1.14 $\sigma$ larger than an extrapolated average
 value for $N$-body simulations, and the latter is another
 0.93 $\sigma$ larger than the collapse time of anisotropic
 gaseous and isotropic Fokker-Planck models. Since this is
 rather large, one may speculate whether it reflects
 a recent result by Takahashi (1995, 1996), who finds
 that in his new 2D FP models collapse times are
 some 10 \% larger than expected from the previous
 isotropic models. However, our particle number is still
 not large enough to discriminate with a sufficient
 statistical significance between such small differences
 in the collapse time.
 \bigskip
 We have also examined the wandering of the density
 centre motions. By autocorrelation analysis and fourier
 transformation we find a dominant period of the order
 of 40-60 time units, which is about 13-20 initial half-mass
 crossing times. Since our database is much longer in time,
 these oscillations could not have been found in previous
 studies like Heggie, Inagaki \& McMillan (1994) and Sweatman (1993).
 The existence of long-time scale oscillations like those we
 found here has, however, been noted already by Makino \& Sugimoto
 (1987).
 \bigskip
 From data of our simulation which were not shown in the plots here
 we conclude that the velocity distribution in the core changes
 significantly between the pre- and post-collapse state of the
 cluster. Thus for example the ratio of the fourth order moment
 of the velocity distribution related to the second order moments
 is larger after core bounce than before, consistent with the idea
 that a high velocity tail of stars is generated by superelastic
 scatterings with the binaries. The presentation of these data
 as well as possible relations to observable line profiles in
 star clusters will be subject of future work. In parallel to
 this it could be useful to extend the gaseous models to higher
 order moments of the velocity distribution in order to achieve
 a better representation of the suprathermal particles in the
 post-collapse phase. One might argue that 2D Fokker-Planck
 simulations, if they become more common in the near future, already
 deliver a complete and unrestricted 2D distribution function. However,
 for the grid-based solutions of the orbit-averaged Fokker-Planck
 equation still an isotropized distribution function is used
 to calculated the diffusion coefficients. This is in contrast
 to the gaseous models, which use a consistent anisotropic background
 for the Fokker-Planck collisional terms. If the Fokker-Planck
 equation is solved by using a variational principle (Inagaki \&
 Lynden-Bell 1990, Takahashi \& Inagaki 1992) it is self-consistent
 (contrary to the discretization method), but there occur other
 uncertainties, as e.g. the proper choice of trial functions.
 We conclude that none of the presently used methods to simulate
 the evolution of large $N$ star clusters should be discarded;
 it will remain very important in the future to produce a larger data base,
 especially for the case of $N$-body simulations with large $N$ and
 to thoroughly study the strong and weak features of each method.
 \bigskip
 To improve the statistics at particle numbers of $N\ge 10000$ more
 independent computations are useful and in progress, especially
 by using the recent fast versions of the HARP special
 purpose hardware (Taiji 1996); to see whether a single star cluster
 behaves in accord with the statistical models and, for example,
 shows gravothermal oscillations, a still larger $N$ is required,
 as in the case of the recent $32000$ particle simulation
 (Makino 1996).
 \par
 \section*{Acknowledgements}
 This work was supported by the {\sl Deutsche
 Forschungsgemeinschaft (DFG)} under grants
 Sp 345/5-1, -2, and 446 JAP 113/18/0.
 The numerical calculations
 were performed at the CRAY Y/MP of the HLRZ.
 The final compilation of this paper
 was done during a stay of R.Sp. at Dept. of Astronomy, Univ.
 of Kyoto, Japan. It is a pleasure for R.Sp. to thank Shogo
 Inagaki and all the colleagues of the department for their
 kind hospitality during this time. R.Sp. is especially grateful
 to Mirek Giersz, Warsaw, Poland, for his hospitality and a
 very fruitful and close cooperation and exchange of data.


\section*{References}
 \beginrefs
 \bibitem Aarseth S.J., 1974, A\&A, 35, 237
 \bibitem Aarseth S.J., 1985, in Brackbill J.U.,
   Cohen B.I., eds, Multiple time scales, Academic Press, Orlando,
      p. 378
 \bibitem Ahmad A., Cohen L., 1973, J. Comput. Phys., 12, 389
 \bibitem Barnes J., Hut P., 1986, Nature, 324, 446
 \bibitem Bettwieser E., Spurzem R., 1986, A\&A, 161, 102
 \bibitem Bettwieser E., Sugimoto D., 1984 MNRAS, 208, 493
 \bibitem Breeden J.L., Cohn H.N., Hut P., 1994, ApJ, 421, 195
 \bibitem Brown, Francis, Rose, Vranesic, 1992, Field
 Programmable Arrays, Kluwer, Dordrecht
 \bibitem Casertano S., Hut P., 1985, ApJ, 298, 80
 \bibitem Cohn H., Hut P., Wise M., 1989, ApJ 342, 814
 \bibitem Dejonghe, H., 1987, MNRAS, 224, 13
 \bibitem Elson, R., Hut, P., Inagaki, S., 1987, ARA\&A, 25, 565
 \bibitem Goodman J., 1987, ApJ, 313, 576
 \bibitem Giersz M., Heggie D.C., 1994a, MNRAS, 268, 257, GHI
 \bibitem Giersz M., Heggie D.C., 1994b, MNRAS, 270, 298, GHII
 \bibitem Giersz M., Spurzem R., 1994, MNRAS, 269, 241, GS
 \bibitem Heggie D.C., Inagaki S., McMillan S.L.W., 1994,
  MNRAS, 271, 706
 \bibitem Hut P., 1993, ApJ, 403, 256
 \bibitem Heggie D.C., 1975, MNRAS, 173, 729
 \bibitem Heggie D.C., 1984, MNRAS, 206, 179
 \bibitem Heggie D.C., Hut P., 1993, APJS, 85, 347
 \bibitem Heggie D.C., Inagaki S., McMillan S.L.W., 1994, MNRAS, 271, 706
 \bibitem Heggie D.C., Mathieu R.M. 1986, in Hut P., McMillan S.L.W.,
  eds, The Use of Supercomputers in Stellar Dynamics. Springer-Verlag,
  Berlin, p. 233
 \bibitem Heggie D.C., Stephenson D., 1988, MNRAS, 230, 223
 \bibitem H\'enon M., 1965, Annales d'Astrophysique, 28, 62
 \bibitem Hills J.G., 1975, AJ, 80, 809
 \bibitem Hut P., Makino J., eds, 1987, Dynamics
 of Star Clusters, Proc. IAU Symp. No. 174, in press
 \bibitem Hut, P., McMillan, S.L.W., eds, The Use of
  Supercomputers in Stellar Dynamics,
  Springer, New York, p. 233
 \bibitem Inagaki S., Lynden-Bell D., 1990, MNRAS, 244, 254
 \bibitem Kustaanheimo P., Stiefel E.L., 1965, J. Reine
  Angew. Math., 218, 204
 \bibitem Louis P.D., Spurzem R., 1991, MNRAS, 244, 478
 \bibitem Lynden-Bell D., Eggleton P.P., 1980, MNRAS, 191, 483
 \bibitem Lynden-Bell D., Wood R., 1968, MNRAS, 138, 495
 \bibitem Makino J., 1996, in Hut P., Makino J., eds, Dynamics
 of Star Clusters, Proc. IAU Symp. No. 174, in press
 \bibitem Makino J., Aarseth S.J., 1992, PASJ, 44, 141
 \bibitem Makino J., Sugimoto D., 1987, PASJ, 39, 589
 \bibitem McMillan S.L.W., Hut P., Makino J., 1990, ApJ, 362, 522
 \bibitem Murphy B.W., Cohn H., Hut P., 1990, MNRAS, 245, 335
 \bibitem Rappaport S., Dewey D., Levine A., Mavri L., 1994,
  ApJ, 423, 633
 \bibitem Spitzer L., 1987, Dynamical Evolution of Globular
  Clusters, Princeton Univ. Press, Princeton
 \bibitem Spurzem R., 1992, in Klare G., ed, Rev. of Modern Astronomy 5,
 Springer Vlg., Berlin, Heidelberg, p. 161
 \bibitem Spurzem R., 1994, in Pfenniger D., Gurzadyan V.G., eds,
 Ergodic Concepts in Stellar Dynamics, Springer-Vlg.,
 Berlin, Heidelberg, p. 170
 \bibitem Spurzem R., 1995, preprint.
 \bibitem Spurzem R., 1996, in Hut P., Makino J., eds, Dynamics
 of Star Clusters, Proc. IAU Symp. No. 174, in press
 \bibitem Spurzem R., Takahashi K., 1995, MNRAS, 272, 772
 \bibitem Sugimoto D., Chikada Y., Makino J., Ito T.,
    Ebisuzaki T., Umemura M., 1990, Nature, 345, 33
 \bibitem Sweatman W.L., 1993, MNRAS, 261, 497
 \bibitem Taiji M., 1996, in Hut P., Makino J., eds, Dynamics
 of Star Clusters, Proc. IAU Symp. No. 174, in press
 \bibitem Takahashi K., 1995, PASJ, 47, 561
 \bibitem Takahashi K., 1996, in Hut P., Makino J., eds, Dynamics
 of Star Clusters, Proc. IAU Symp. No. 174, in press
 \bibitem Takahashi K., Inagaki S., 1992, PASJ, 44, 623
 \endrefs
 \bye